\documentclass{article}
\usepackage{arxiv}
\usepackage{amssymb}
\usepackage{amsmath}
\usepackage{graphicx}
\usepackage{makeidx}
\usepackage{multicol}
\usepackage{listings}
\usepackage{pdflscape}
\usepackage{tabu}
\usepackage{makecell}
\usepackage{textcomp}
\usepackage{subfig}
\setcellgapes{3pt}

\title{Swarm Intelligence for Morphogenetic Engineering}
\author{
Bruce J.\ MacLennan\\
Dept.\ of Electrical Engineering and Computer Science\\
University of Tennessee, Knoxville \\
\texttt{maclennan@utk.edu}\\
\And
Allen C.\ McBride\\
Dept.\ of Electrical Engineering and Computer Science\\
University of Tennessee, Knoxville\\
\texttt{amcbri10@vols.utk.edu}\\
}

\begin{document}
\maketitle

\begin{abstract}
We argue that embryological morphogenesis provides a model of how massive swarms of microscopic agents can be coordinated to assemble complex, multiscale hierarchical structures.
This is accomplished by understanding natural morphogenetic processes in mathematical terms, abstracting from the biological specifics, and implementing these mathematical principles in artificial systems.
We have developed a notation based on partial differential equations for artificial morphogenesis and have designed a prototype morphogenetic programming language, which permits precise description of morphogenetic algorithms and their automatic translation to simulation software.
Morphogenetic programming is illustrated by two examples:
(1) use of a modified flocking algorithm to route dense fiber bundles between regions of an artificial cortex while avoiding other bundles;
(2) use of the clock-and-wavefront model of spinal segmentation for the assembly of the segmented spine of an insect-like robot body and for  assembling segmented legs on the robot's spine.
Finally, we show how a  variation of smoothed particle hydrodynamics (SPH) swarm robotic control can be applied to the global-to-local compilation problem, that is, the derivation of individual agent control from global PDE specifications.

\end{abstract}


\newcommand{\PreChap}[2]{#1}
\newcommand{\parencite}[1]{\cite{#1}}
\newcommand{\textcite}[1]{reference \cite{#1}}

\def\DH{\mbox{\raisebox{0.2ex}{{\rm --}}}\hspace{-0.88ex}{\rm D}}
%
%
\def\figdir{figures/}
%
%
\newcommand{\Fig}[3]{
\begin{figure}[b!]
\centerline{\includegraphics[width=150pt, height=150pt]{\figdir #2}}
\caption{#3}
\label{fig:#1}
\end{figure}
}
%
%
\newcommand{\GenFig}[4]{
\begin{figure}[b!]
\centerline{\includegraphics[#3]{\figdir #2}}
\caption{#4}
\label{fig:#1}
\end{figure}
}
%
%
\newcommand{\MultiFig}[3]{
\begin{figure}
\begin{center}
#2
\end{center}
\caption{#3}
\label{fig:#1}
\end{figure}
}
%
%
\newcommand{\SubFig}[4]{
\subfloat[#4\label{fig:#1}]{\includegraphics[#3]{\figdir #2}}
}
%
%
\newcommand{\Citepar}[1]{\parencite{#1}}
%
%
%
\def\d{{\rm d}}				
\def\D{{\rm D}}				
\def\ETH{\DH}				
\def\into{\rightarrow}			
%
%
\def\Change{\ETH}				
\def\convolve{\otimes}			
\def\del{\nabla}					
\def\vdel{\boldsymbol{\del}}		
\def\Deltat{\Delta_t}				
\def\Dt{\D_t}					
\def\distributed{\sim}				
\newcommand{\evaledAt}[2]{\left. #1 \right|_{#2}}
\def\fixedElem{\anElem\,\mbox{\scriptsize fixed}}
\def\fixedElemPos{\anElemPos\,\mbox{\scriptsize fixed}}
\def\Lapl{\del^2}				
\newcommand{\norm}[1]{\| #1 \|}	
\def\regBy{\sim} 				
\newcommand{\regulation}[2]{#1 \regBy #2}	
\def\trans{^{\rm T}}				
\def\unitstep{u}					
\def\incrby{\;+\!\!=}				
\def\decrby{\;-\!\!=}				
\newcommand{\If}[1]{[ #1 ]}		
%
%
\def\EuclideanSp{{\cal E}}			
\def\aBody{{\cal B}}				
\newcommand{\Reals}{\mathbb{R}}	
\def\aRegion{{\cal R}}			
%
%
\newcommand{\unit}[1]{\vect{e}_#1}	
\def\Activator{A}				
\def\anElem{P}					
\def\anElemvec{\vect{\anElem}}		
\def\anElemPos{\vect{p}}			
\def\aPropS{q}					
\def\aPropM{Q}					
\def\anElemVelCom{v}			
\def\anElemVel{\vect{\anElemVelCom}}	
\def\anElemAcc{\vect{A}}			
\def\driftvec{\bm{\mu}}		
\def\difftens{\mbox{\boldmath $\sigma$}}	
\def\Dtensor{\vect{D}}				
\newcommand{\freeVariable}[1]{\delta #1} 
\def\freeA{\freeVariable{A}}	
\def\freeL{\freeVariable{L}}	
\def\freeV{\freeVariable{V}}	
\def\Inhibitor{I}				
\def\Knudsen{{\rm Kn}}		
\def\noise{\nu}				
\def\numDen{n}				
\def\probDen{\phi}			
\def\Reynolds{{\rm Re}}		
\def\uvec{\vect{u}}			
\def\Uvec{\vect{U}}			
\def\vvec{\vect{v}}			
\def\Vvec{\vect{V}}			
\def\Wiener{W}				
\newcommand{\Wienervec}[1]{{\bf W}^{#1}}	
%
%
\def\cement{C}				
\def\pheromone{\phi}		
\def\terminflux{r_{\termites}}	
\def\termites{A}				
%
%
\newcommand{\DecayRate}[1]{\tau_#1}	
\newcommand{\DiffRate}[1]{D_#1}	
\newcommand{\Const}[1]{\kappa_#1}	
\newcommand{\Constant}[1]{c_#1}	
\newcommand{\UPB}[1]{#1\subupb}	
\newcommand{\LWB}[1]{#1_{\rm lwb}}	
\newcommand{\Threshold}[1]{\vartheta_#1}	
\def\Anterior{R}				
\def\Posterior{C}			
\def\Smorphogen{\alpha}		
\def\Tail{T}				
\def\Medium{M}				
\def\Somites{S}				
\def\SomiteRate{\kappa_\Somites}	
\def\SomiteTrigger{\varsigma}	
\def\subupb{_{\rm upb}}		
\def\AntUpb{\Anterior\subupb}	
\def\PstUpb{\Posterior\subupb}	
\def\AntDiff{D_\Anterior}		
\def\PstDiff{D_\Posterior}		
\def\SDiff{D_\Smorphogen}	
\def\AntDecay{\tau_\Anterior}	
\def\PstDecay{\tau_\Posterior}	
\def\SDecay{\tau_\Smorphogen}	
\def\AntSat{\kappa_\Anterior}	
\def\PstSat{\kappa_\Posterior}	
\def\Clock{K}				
\def\Clockfreq{\omega}		
\def\ClockfreqHz{\nu}                
\def\Clockthresh{\vartheta_\Clock}	
\def\ClockCond{\psi}			
\def\Sthresh{\vartheta_\Smorphogen}	
\def\SLwb{\Smorphogen_{\rm lwb}}	
\def\Rec{\varrho}			
\def\RecDecay{\tau_\Rec}		
\def\Recthresh{\vartheta_\Rec}	
\def\FireCond{\phi}			
\def\FireRate{\kappa_\FireCond}	
\def\TailDirection{\uvec}		
\def\TailRate{r}				
\def\TailLength{\lambda}		
\def\SegGap{g}                         
\def\GrowthTimer{G}			
\def\GrowthDecay{\tau_\GrowthTimer}	
\def\Growththresh{\vartheta_\GrowthTimer}	
\def\GrowthDuration{t_\GrowthTimer}	
\def\GrowthAttractant{H}            
\def\GrowthRate{v}                     
\def\Source{J}                             
\def\BodyLength{\Length{{\rm B}}}	
\def\AntBlock{B}                        
\def\AntBorder{A}			
\def\PostBorder{P}			
\def\AntBordMorph{a}		
\def\PostBordMorph{p}		
\def\GaussKernel{\gamma}	
\def\SmoothDensity{N}              
\def\Edge{E}				
\def\ImagDisk{I}			
\def\IDchangethresh{\Threshold{{\rm D\ImagDisk}}}	
\def\LegInit{\zeta}                      
\def\LegTissue{L}		        
\def\LegMorph{\mu}                   
\def\LegBlock{Z}                        
\def\LegBlockMorph{z}              
\def\App{{\rm A}}			
\def\LegLength{\Length{\App}}	
\def\SpineSegLength{\lambda_{\rm S}}	
\def\LegSegLength{\lambda_{\rm L}}	
\def\SpineSegNum{N_{\rm S}}  
\def\LegSegNum{N_{\rm L}}     
%
%
\def\attractant{A}			
\def\Adiff{D_\attractant}		
\def\Adecay{\tau_\attractant}	
\def\Aaccel{\gamma}			
\def\goal{G}				
\def\goalrate{\kappa_\goal}	
\def\path{P}				
\def\pathdep{\beta}			
\def\clamping{\tau_\path}		
\def\cones{C}				
\def\conerate{\alpha}			
\def\conediff{\sigma}			
\def\conedensmorph{c}		
\def\ranvec{\vect{\Wiener}}	
\def\source{\vect{s}}			
\def\dest{\vect{g}}			
\newcommand{\urge}[1]{\vect{u}_{\rm #1}} 
\newcommand{\uwgt}[1]{w_{\rm #1}} 
\def\swarmrate{\kappa_\cones}	
\def\swarmversor{\vect{v}}		
\def\gradstrength{S}			
\def\gradthreshold{\theta}		
\def\densweight{\lambda}		
\def\dpotential{U}			
\def\potential{W}			
%
%
\newcommand{\Normal}[2]{\mathcal{N}(#1, #2)}	
\def\Config{C}				
%
\newcommand{\Conf}[2]{C_#1(#2)}
\newcommand{\Confinv}[2]{C^{-1}_#1(#2)}	
\newcommand{\Prob}[1]{\mathop{\rm Pr}\{#1\}}
%
%
\newcommand{\PL}[1]{{\bf #1}}	
%
%
%
\newcommand{\SubstanceDecls}[1]{
\begin{tabbing}
MMMM\=order-9\= fields MMM \= //comments\kill
#1
\end{tabbing}
}
%
%
\newcommand{\Substance}[3]{
\substance #1#2:
\vspace*{-0.1in}
\begin{tabbing}
MM\=order-9\= random MMM \= //comments\kill
#3
\end{tabbing}
}
%
%
\newcommand{\BehaviorDefs}[1]{
\hspace*{0.2in}\behavior:
\begin{eqnarray*} #1 \end{eqnarray*}}	
\def\commark{$\|$}						
%
%
\newcommand{\Srem}[1]{{\ \ \ \  \commark\ #1}}	
%
%
\newcommand{\Brem}[1]{\mbox{\ \ \ \  \commark\ #1}}	
%
%
\newcommand{\BremOnly}[1]{\lefteqn{ \mbox{\commark\ #1}}\\ }
%
%
\def\substance{\PL{substance\ }}
\def\is{\PL{is\ }}
\def\with{\PL{with}}
\def\scalar{\PL{scalar\ }}
\def\scalars{\PL{scalars}}
\def\vector{\PL{vector\ }}
\def\order{\PL{order}}
\def\field{\PL{field}}
\def\fields{\PL{fields}}
\def\random{\PL{random}}
\def\behavior{\PL{behavior}}
%
%
\newcommand{\TypeVar}[3]{\>#1\ $#2$\>\>\Srem{#3}\\}	
%
%
\newcommand{\Type}[1]{\>#1:\\}			
%
%
\newcommand{\Var}[2]{\>\>$#1$ \>\Srem{#2}\\}	
%
%
\def\body{\PL{body\ }}
\def\of{\PL{of\ }}
\def\for{\PL{for\ }}
\def\And{\PL{and}}
%
%
\newcommand{\BodyHead}[2]{
\body #1 \of #2:\\
}
%
%
\def\BodyTail{
}
%
%
\newcommand{\ForAll}[1]{
#1 
}
%
%
\newcommand{\For}[2]{
\hspace*{0.52in}\for $#1$:\\
#2 
}
%
%
\newcommand{\ForOne}[2]{
\hspace*{0.52in}\for $#1$:
$#2$ \\
}
%
%
\newcommand{\Init}[2]{\hspace*{0.75in}$#1$ = $#2$\\ }
%
%
\newcommand{\Body}[4]{
\BodyHead{#1}{#2}
\For{#3}{#4}
\BodyTail
}
%
%
%
\newcommand{\Section}[2]{\section{#1}\label{secnum:#2}} 
\newcommand{\Subsection}[2]{\subsection{#1}\label{secnum:#2}} 
\newcommand{\Subsubsection}[2]{\subsubsection{#1}\label{secnum:#2}} 
\newcommand{\CSubsubsection}[2]{\noindent\par{\bf #1:}\label{secnum:#2}} 
\newcommand{\secref}[1]{Section \ref{sec:#1} \secname:#1}	
\newcommand{\Secref}[1]{Sec.\ \ref{sec:#1}}		
\newcommand{\Citem}[2]{\noindent\par {\bf (#1) #2:}}		
%
%
\def\Ito{It\={o}}				
\newcommand{\OMIT}[1]{}	
\newcommand{\NB}[1]{\begin{quote}\large\bf #1\end{quote}}	

\hyphenation{na-no na-no-tech-no-lo-gy na-no-struc-ture na-no-struc-tures}
\hyphenation{na-no-com-pu-ta-tion na-no-tube na-no-tubes na-no-scale}
\hyphenation{na-no-struc-tur-al}
\hyphenation{morph-gen}
\lstdefinelanguage{morphgen}
	{morekeywords={morphogenetic,program,end,
	simulation,parameters,space,
	spatial,temporal,resolution,duration,
	substance,behavior,scalar,vector,field,fields,
	D,let,param,params,
	del,div,DW,
	body,of,for,within,
	report,diffusion,Courant,Peclet,number,and,
	visualization,interval,display,final,running,as,mesh,colors,contours,quivers,grid,limits,code,
	make,movie,
	save,to,load,from,
	},
	morecomment=[l]{//},
	morecomment=[s]{/*}{*/},
	stringstyle=\ttfamily,
	morestring=[b]",}
\lstset{language=morphgen}
\newcommand{\word}[1]{{\bf #1}}
\newcommand{\nterm}[1]{\langle\mbox{#1}\rangle}
\def\nl{\nterm{newline}}
\def\id{\nterm{indent}}
\def\dd{\nterm{dedent}}
\def\caret{\pmb{\hat{\ }}}
\def\uscore{\pmb{\_}}
\def\bra{\pmb{[}}
\def\ket{\pmb{]}}
\newcommand{\neqn}[2]{\begin{equation}\label{eq:#1}
#2
\end{equation}}
\def\Morphgen{Morphgen}         
\def\MORPHGEN{Morphgen}         
\newcommand{\vect}[1]{{\bf #1}}	
\def\DW{\Change\Wiener}
\def\Uvec{{\bf U}}
\def\Vvec{{\bf V}}
\def\Wvec{{\bf W}}
\def\divergence{\mathop{\rm div}}	

\section{Self-assembly of Complex Hierarchical Structures}
We are applying swarm intelligence to the coordination of microrobot swarms to assemble complex, hierarchically structured physical systems.
We are interested in multiscale systems that are intricately and specifically structured from the microscopic level up through the macroscopic level: from microns to meters.
On one hand, conventional manufacturing techniques, including additive manufacturing, work well at the macroscopic level, but are difficult to scale down below the millimeter scale.
On the other, technologies that are effective at the microscopic scale, such as photolithography and molecular self-assembly, do not scale well to complex macroscopic objects.

Nevertheless, it would be valuable to be able to assemble automatically systems structured across many length scales.
For example, the human brain has complex structures spanning more that six orders of magnitude, from micron-scale synapses to decimeter-scale functional regions and interconnections 
\Citepar{Lichtman-Denk}.
It is not unreasonable to suppose that a neuromorphic computer with capacities and functions similar to a human brain would have similar complexity.
How would we assemble a neuromorphic computer with 100 billion neurons and perhaps 100 trillion nonrandom connections? 
Similarly complex future sensors and actuators, with capabilities similar to animal sense organs and effectors, will span scales from the microscopic to the macroscopic.

\par 
Automatic assembly of such complex structures might seem to be an unobtainable goal, but we know it is possible, for nature does it.
Embryos develop from a single cell to a complex organism with many trillions of cells, each itself a complex hierarchical system.
During development, cells communicate, coordinate, and cooperate, behaving as a massive swarm, to differentiate and rearrange into the various interrelated tissues that constitute a complete organism.
\emph{Morphogenesis} refers to the developmental process that assembles and organizes three-dimensional forms and structures.
The process is massively parallel, distributed, and robust: all goals for us too.

\par 
\emph{Morphogenetic engineering} or \emph{artificial morphogenesis}  takes inspiration from biological morphogenesis, seeking new ways to create complex structures that share some ideas from the development of embryos in multicellular organisms. 
Because of the diversity of mechanisms in biological morphogenesis from which inspiration might be taken, morphogenetic engineering is diverse.
It can overlap several fields including swarm robotics, modular robotics, amorphous computing, cellular automata, and synthetic biology.
Bodies developed through morphogenetic engineering may take form through the birth and death of sessile elements, or through the rearrangement of motile ones.
Design approaches also vary.
Processes may be designed by hand in an ad hoc fashion to illustrate particular hypotheses or principles.
Designs may also be generated by hand in more systematic ways, for example through global-to-local compilation.
Evolutionary algorithms are also popular for automatic design of processes to achieve well-defined goals.
Finally, a spectrum exists from biological morphogenesis with limited human control to fully artificial systems 
\parencite{teague_synthetic_2016}.

Embryological morphogenesis has inspired a variety of approaches to swarm intelligence 
\Citepar{%
Kitano-MES,%
Nagpal-Kondacs-Chang,%
Murata-Kurokawa-SRR,%
Spicher-Michel-Giavitto-ASA,%
Dourset-OGA,%
Bourgine-Lesne-MOPS,%
Giavitto-Spicher-CM%
};
our approach to artificial morphogenesis tends to adhere more closely to natural morphogenesis than do most of these others.
\textcite{doursat_review_2013} provide a rigorous taxonomy of morphogenetic engineering,
including additional examples, 
and \textcite{oh_bio-inspired_2017} provide a more recent review of similar work.

Morphogenetic engineering is a specific approach to the goal of \emph{programmable matter}, that is, the ability to systematically control the properties and behavior of material systems at a fine level
\Citepar{Goldstein-PM,MPPM},
and artificial morphogenesis exhibits the properties of \emph{active matter}
\Citepar{Needleman-Dogic-AM,Doostmohammadi-AN,Xi-Saw-MAATM}.
Morphogenetic engineering also has some similarities to \emph{amorphous computing} 
\Citepar{Abelson-AC},
but artificial morphogenesis processes may begin with simple, structured preparations (see for example \Secref{segmentation} below), and even if they do begin in an unorganized state, it is characteristic of morphogenetic processes to quickly self-organize.

\par 
Although natural morphogenesis is a complex and intricate process---still incompletely understood---biologists have identified about twenty fundamental processes
\cite[pp.\ 158--9]{F-N}\cite{Salazar-Ciudad-et-al}.
Not all are applicable to artificial morphogenesis, and some (such as cell division) may be difficult to implement, nevertheless these processes provide an agenda for morphogenetic engineering, since they are in principle sufficient for assembling something as complex as an animal body
\Citepar{PDFECAM,MMNC,MCHSA,CMRS}.
Processes that cannot be directly implemented (such as cell division) may need to have their function accomplished by alternative means (e.g., providing components from an external source and moving them to the growth zone; see \Secref{growth process} for an example).
   
\par 
During biological morphogenesis, cells emit and receive chemical signals, migrate and differentiate in response to those signals, and participate in both reproduction (\emph{cell proliferation}) and programmed cell death (\emph{apoptosis}).
In the process they emit and absorb molecules that serve both as communication media (\emph{morphogens}) and as structural elements.
Similarly, in our approach to artificial morphogenesis we distinguish \emph{active components} and \emph{passive components} \Citepar{MPPM}.
Active components are microscopic agents (microrobots or genetically engineered micro-organisms) that can emit and respond to simple signals, implement simple (primarily analog) control processes, and move and transport other (active or passive) components.
Passive components are all the rest, including signaling molecules and structural components.
They do not move under their own power, but are moved by external forces (including Brownian motion and active components).
In a typical artificial morphogenesis process, agents (active components) might transport and assemble passive components into a desired structure, or they might assemble themselves into the structure, as cells do in biological development.

\section{Describing Artificial Morphogenesis with PDEs}
The challenge of artificial morphogenesis is to control very large swarms of active components (microscopic agents) to interact with each other and with the passive components to assemble a desired structure.
This requires an appropriate level of abstraction that allows the process to be described in sufficient detail for implementation without becoming obscured by details of a specific implementation technology.
Here again we have taken our inspiration from biology, for biologists often use partial differential equations (PDEs) to describe morphogenetic processes.
At this level of abstraction developing tissues and diffusing signaling chemicals are described at a macroscopic level largely independent of individual cells.
It also allows the methods of continuum mechanics to be applied, which are especially appropriate for embryological development, which takes place in the domain of ``soft matter'' (viscoelastic materials)
\Citepar{deGennes-SM,F-N,Taber-NTE}.
Therefore we are operating in the domain of well-understood mathematical methods in which continuum models are applied to macroscopic volumes of materials with a discrete molecular structure.

Scalability is a principal advantage of using PDEs for morphogenetic engineering.
This is because a PDE, especially when used in continuum mechanics or fluid dynamics, treats a material as a phenomenological continuum; that is, it is treated as though composed of an infinite number of infinitesimal particles, which is a usable approximation of a material composed of a very large number of very small particles.
Therefore, if our goal in artificial morphogenesis is to use a very large swarm of very small agents, PDEs take this goal to the continuum limit.
Instead of worrying whether our algorithms will scale up to larger swarms of smaller agents, we have the complementary, but easier, task of scaling down from an infinite number of infinitesimal agents to a very large number of very small agents.

Another advantage of expressing morphogenetic processes in PDEs is that they are largely independent of agent size.
This is because the processes are expressed in terms of \emph{intensive quantities}, such as agent density, rather than \emph{extensive quantities}, such as numbers of agents \Citepar{PDFECAM,MMNC,CSMA}.
Therefore, an algorithm that produces structures of a particular size will continue to be correct even if the size of the agents is changed.
(There are obviously limits to this size independence; the continuum approximation has to be good.)

In summary, using PDEs allows us to describe the behavior of \emph{massive swarms}, by which we mean swarms that can be treated as a continuous mass.
We treat massive swarms the same way tissues are treated in embryological morphogenesis and the same way fluids and solid masses are treated in continuum mechanics.

\section{A Morphogenetic Programming Language}
In order to test algorithms for the coordination of microrobot swarms for morphogenetic engineering, we have developed a morphogenetic programming notation based on mathematical notation for PDEs
\Citepar{PDFECAM,MMAM,AMEEC,EC-APCAM,MCHSA}.
In order to facilitate simulations, this notation has been formalized into a morphogenetic programming language, tentatively named ``\Morphgen .''
We have a prototype implementation of \Morphgen\  by means of a syntax macroprocessor, which translates \Morphgen\  programs into {\sc Matlab}\textsuperscript{\textregistered} or compatible GNU Octave \Citepar{SSM-TR5}.
This approach imposes some syntactic limitations on the \Morphgen\  language, which would not be required with a conventional compiled implementation, but it permits rapid prototyping and experimentation with language features.
There are two slightly different dialects of the language for describing either two-dimensional or three-dimensional morphogenetic systems.
(A complete grammar for a previous version of \Morphgen\  is published in \textcite{MPPFCF};
here we use \Morphgen\ version 0.13.)
\subsection{Substances}
A morphogenetic program is organized into a number of \emph{substances} with common properties;
typical substances include diffusing morphogens and massive swarms of agents.
Substances are similar to classes in object-oriented programming languages, in that a substance defines the common properties of an unlimited number of particular instances (called \emph{bodies} in \Morphgen , analogous to objects in object-oriented programming).
Like subclasses in object-oriented programming, more specific substances may be derived from more general substances by specification of parameters that were unspecified in the more general substance, and by the addition of properties and behaviors in the derived substance.
For example, we might have a substance representing a general diffusible substance with an unspecified diffusion rate, and one or more specific diffusible substances, with specified diffusion rates, derived from the general substance.

We distinguish \emph{physical substances} and \emph{controllable substances}, but this is more a matter of degree than kind, and applies primarily to active components.
The idea is that the properties and behavior of physical substances are relatively fixed.
For example, a particular chemical morphogen will have specific diffusion and decay rates within a given medium,
and a particular agent swarm will be composed of agents, such as microrobots, with particular masses and specific sensors and actuators.
Changing these properties entails making a new substance, new kinds of agents, and different hardware.

In contrast, a controllable substance has properties that are relatively easy to control.
For example, a programmable agent will permit its sensors and actuators to be controlled so that it behaves in a desired way; its software can be changed.
That is, a controllable substance is in some sense programmable, but there are degrees of programmability.

The current \Morphgen\  language does not distinguish between physical and controllable substances, since controllability is a matter of degree and even economics.
However, the distinction might be expressed in a substance hierarchy.
For example, a substance definition might reflect the physical properties and behavior of a certain kind of microrobot, and then various substances derived from it could represent the microrobots programmed for distinct functions and behaviors.

\par 
This is an example of a simple substance definition (a diffusible morphogen), which illustrates its parts:
\newpage 
\label{eg-substance}
\begin{lstlisting}[frame=single]
substance morphogen:
    scalar field C      // concentration
  behavior:
    param d_C = 0.3     // diffusion constant
    param t_C = 10      // decay time constant
    D C = d_C * del^2 C - C/t_C // diffusion & decay
\end{lstlisting}
As illustrated by this example, the syntactic extent of many \Morphgen\  constructs is indicated by indenting.
(This substance definition is terminated by a line indented less or equal that of the word $\word{substance}$.)
A substance definition has two parts: field declarations (preceding $\word{behavior}$) and behavioral specifications (following $\word{behavior}$).

\subsection{Field Declaration}
A substance is characterized by one or more continuous \emph{fields} defined throughout the space.
These may be scalar fields, such as concentration or density fields,
or vector fields, such as velocity or flux fields.
The preceding definition of ``morphogen'' illustrates the declaration of a scalar field called ``C'';
an example vector field declaration is:
\begin{lstlisting}
    vector field V
\end{lstlisting}
Multiple fields of the same type can be declared on indented lines, as illustrated by this example:
\begin{lstlisting}
    scalar fields:
      C    // swarm density
      S    // magnitude of morphogen gradient
    vector fields:
      U    // morphogen gradient
      V    // swarm velocity
\end{lstlisting}

\subsection{Behavior Specification}
Fundamental to our approach is the use of partial differential equations to describe the behavior of massive swarms of microscopic agents.
Such agents and the materials that they control will move continuously in time, but we also simulate this behavior on ordinary computers in discrete time.
Therefore we describe the behavior of substances by \emph{change equations} which can be interpreted ambiguously as either ordinary PDEs or as temporal finite difference equations.
The derivation rules of this calculus respect both interpretations.
We write change equations with the notation
$\Change X = F(X,Y,\ldots)$, which means either
the differential equation $\partial_t X = F(X,Y,\ldots)$ 
or the difference equation $\Delta X / \Delta t = F(X,Y,\ldots)$.
A change equation such as
\[ \Change C = d_C \del^2 C - C/t_C \]
is written as follows in the \Morphgen\  language:
\begin{lstlisting}
    D C = d_C * del^2 C - C/t_C 
\end{lstlisting}
\par 
The \Morphgen\  prototype implementation does not include a full expression parser, and so there are some notational concessions to allow expressions to be handled by the syntax macroprocessor.
For the most part, scalars and fields (both scalar and vector) can be combined using standard arithmetic operators (+, \lstinline'-', \lstinline'*', /).
However, products or quotients of two scalar fields, products of a scalar field and a vector field, quotients of a vector field and a scalar field, and powers of scalar fields must be surrounded by square brackets,
for example, \lstinline'[C * V]'.
Morphgen includes several vector operators, including
the gradient \lstinline'del X',
the Laplacian \lstinline'del^2 X',
the divergence \lstinline'div X',
and the pointwise $\mathcal{L}_2$ norm of a vector field, \lstinline'||X||'.
\par 
Some change equations are stochastic, either to model indeterminacy and uncertainty in the physical systems, or to introduce randomness into the swarm control (e.g., to break deadlocks and symmetry).
In order to respect both continuous- and discrete-time stochastic change equations, the morphogenetic programming notation interprets $\Change\Wiener^n$ to be an $n$-dimensional random vector field distributed $\Normal{0}{1}$ in each dimension; it represents $n$ independent sources of randomness at each point in the (two- or three-dimensional) space
\Citepar{PDFECAM,MMAM,AMEEC}. 
For a $d \in \{2, 3\}$ dimensional morphogenetic process, we usually use $M\Change\Wiener^n$, where $M$ is a $d \times n$ matrix that weights and sums the random inputs into a $d$-dimensional vector field.
$M$ can be a $1 \times n$ matrix to generate a scalar field, or a scalar when $n \in \{1, d\}$.
All of these cases are written \lstinline'[M DW^n]' in the \Morphgen\  language.

\par 
Control of morphogenetic swarms often requires them to change their behavior when some quantity exceeds a threshold or is within some range.
To accommodate these requirements, \Morphgen\  permits bracketed conditional factors (actually, Heaviside step functions), which are 0 or 1 depending on whether or not the condition is true.
For example, the following describes a scalar field $A$ in which any value greater than $\theta$ increases exponentially until it saturates at $A=1$; values less than or equal to $\theta$ decay to zero (with time constant $\tau$):
\begin{lstlisting}
    D A = [A > theta] A * (1-A) - [A <= theta] A/tau
\end{lstlisting}

\par 
Sometimes the behavior of a field is described by several \emph{partial change equations} distributed among several substance definitions.
Partial change equations are written in either of these forms:
\begin{eqnarray*}
\word{D}\ \nterm{name}\ \incrby \nterm{expr}\nl\\
\word{D}\ \nterm{name}\ \decrby \nterm{expr}\nl
\end{eqnarray*}
(This notation was first used in morphogenetic programming by \textcite[p.\ 20]{Fleischer-PhD}.)
A typical application would be describing the behavior of a morphogen.
The definition of the morphogen substance might define its diffusion and decay with a partial change equation such as this:
\begin{lstlisting}
    D M += d_M * del^2 M - M / tau_C 
\end{lstlisting}
This is part of the definition of the morphogen because it is a physical property of the substance.
Elsewhere in the program, in the definition of a substance representing the agent swarm (with swarm density $S$), another partial change equation could describe the emission of the morphogen at a rate $k_M$:
\begin{lstlisting}
    D M += k_M * S 
\end{lstlisting}
This is part of the behavior of the swarm substance because the morphogen is produced by the agents constituting the swarm.
\par 
It is often convenient to name additional scalars or fields within a substance behavior definition, which is accomplished by a statement of the form:
\[ \word{let}\ \nterm{name} = \nterm{expr}\nl \]
(In the current prototype implementation, declared fields have global scope, but $\word{let}$-defined variables are local to the substances in which they are defined.)
\par 
The behavior of substances is determined also by certain fixed parameters, such as diffusion and decay rates. These are specified by a \emph{parameter definition}, which has the syntax:
\[ \word{param}\ \nterm{name} = \nterm{expr}\nl \]
Parameter definitions usually occur at the beginning of the $\word{behavior}$ part of a substance definition,
but can also be global as part of the $\word{simulation}$ $\word{parameters}$ 
(see \Secref{simulation} below).
A sequence of parameter definitions can be indented under the word $\word{params}$.

The following behavioral rules \Citepar{MPPFCF} illustrate many of the \Morphgen\  features we have discussed
(and also line continuation indicated by ``\lstinline"..."''):
\begin{lstlisting}[frame=single]
    params:
      v = 1          // base swarm speed
      lambda = 0.03  // density regulation
      eps = 1e-100   // minimum gradient norm
      k_W = 0.1      // degree of random motion
      k_P = 30       // path deposition rate
      t_D = 5        // delay period
    let U = del A
    let S = ||U||
    let V = [(v*U)/(S+eps)] - lambda*del[(C-1)^2] ...
      + [k_W DW^2]
    D C = [t>t_D] -div[C*V]  // change in density
    D P += [t>t_D] k_P*[C*(1-P)] // path deposition
\end{lstlisting}

\subsection{Bodies}\label{sec:Bodies}
The substance definitions of a morphogenetic program describe the general properties of the materials or substances (including massive agent swarms) that are involved in the process.
A complete morphogenetic program also requires specification of the initial conditions, which in the case of morphogenesis means an initial preparation of the substances.
Since the goal of morphogenesis is for complex structure to emerge by means of self-assembly, the initial preparation should be simple in structure: simple spatial arrangements of materials and of sources and sinks.

In our case, the initial conditions are specified by defining a small number of \emph{bodies} belonging to defined substances.
A body definition has a header followed by a sequence of (indented) initializations, for example:
\begin{lstlisting}
body Obstacles of path_material: // place obstacles
  for (x, y) within 0.06 of (-0.1, 0.225): P = 1
  for (x, y) within 0.06 of (0.1, -0.225): P = 1
\end{lstlisting}
Initializations currently have the following forms:
\begin{eqnarray*}
\nterm{initialization} &::=& \word{for}\ \nterm{region}: \nterm{init}\\
  &|& \word{for}\ \nterm{region}:\nl\ \nterm{init}^*\ \dd\\
\nterm{region} &::=&
  \nterm{expr} < \nterm{name} < \nterm{expr}, \nterm{expr} < \nterm{name} < \nterm{expr}\\
  &|& (\nterm{name}, \nterm{name})\ \word{within}\ \nterm{expr}\ \word{of}\ (\nterm{expr}, \nterm{expr}) \\
\nterm{init} &::=& \nterm{name} = \nterm{expr}\nl
\end{eqnarray*}
($\dd$ represents delimitation by an indent level less than or equal to the the beginning of the construct.)
This syntax is for the 2D version of \Morphgen ; the 3D version has similar initializations.
The second kind of $\word{for}$ definition permits multiple fields to be initialized in a region, e.g.:
\begin{lstlisting}
body Start of swarm
  for -0.5<x<0.5, 0<y<0.1:
    S = 1     // initial swarm density
    M = 0.05  // initial morphogen concentration
\end{lstlisting}
(Uninitialized regions are assumed to be zero.)
\OMIT{
\subsection{Boundary Conditions} \fbox{OMIT?}
     3.6.1, Space.
     3.6.2, Dirichlet.
     3.6.3, Neumann.
     3.6.4, Cauchy.
}
\subsection{Simulation}\label{sec:simulation}
The preceding morphogenetic programming constructs apply as well to simulations as to specifying real morphogenetic processes executed by physical microrobots.
For simulation purposes, the \Morphgen\  language provides additional facilities.
The overall structure of a \Morphgen\  program has the syntax:
\begin{eqnarray*}
 && \word{morphogenetic\ program}\ \nterm{name}:\\
 &&\ \ \ \ \nterm{sim params}\\
 &&\ \ \ \ \nterm{substance}^*\\
 &&\ \ \ \ \nterm{body}^*\\
 &&\ \ \ \ \nterm{visualization}\\
 &&\word{end\ program}
\end{eqnarray*}
The parameters for the simulation are followed by the substance definitions, which are followed by the body definitions, which are followed by visualization commands.

The block of simulation parameters have the syntax:
\begin{eqnarray*}
\nterm{sim params} &::=& \word{simulation\ parameters}: \nterm{sim par}^*\ \dd \\
\nterm{sim par} &::=& \word{duration} = \nterm{num}\nl\\
  &|& \word{temporal\ resolution} = \nterm{num}\nl\\
  &|& \word{space}\ \nterm{num} < \word{x} < \nterm{num}, \nterm{num} < \word{y} < \nterm{num} \nl\\
  &|& \word{spatial\ resolution} = \nterm{num}\nl\\
  &|& \word{save}\ \nterm{name}^+\ \word{to}\ \nterm{filename}\nl\\
  &|& \word{load}\ \nterm{name}^+\ \word{from}\ \nterm{filename}\nl\\
  &|& \nterm{log params}\\
\nterm{log params} &::=& \word{log\ params}\ \nterm{name}\ [,\ \nterm{name}]^*\ \nl\\
  &|& \word{log\ note}\ \nterm{characters}\ \nl
\end{eqnarray*}
The $\word{space}$ specification defines the (two- or three-dimensional) region in which the morphogenetic process takes place; $\word{duration}$ determines how long (in simulated time) the simulation runs.
The spatial and temporal resolution of the simulation are defined by the $\word{spatial\ resolution}$ and $\word{temporal\ resolution}$ specifications.
The $\word{save}$ and $ \word{load}$ directives allow scalar and vector fields to be saved and restored.
Finally, $\word{log\ params}$ records the specified parameter values in a time-stamped text file,
and $\word{log\ note}$ puts a note in it.

Morphgen provides visualization commands to display fields in a variety of formats either during the simulation ($\word{running}$) or at its end ($\word{final}$):
\begin{eqnarray*}
\nterm{visualization} &::=& \word{visualization}: \nterm{command}^+\ \dd\\
\nterm{command} &::=&
  \word{display}\ \nterm{time}\ \nterm{primitive}\ \word{as}\ \nterm{kind}\ \nterm{options}\\
  &|& \word{make}\ \word{movie}\ \nterm{filename}\ \word{of}\ \nterm{primitive}\ \word{as}\ \nterm{kind} \ \nterm{options}\\
  &|& \nterm{stability report}\\
\nterm{time} &::=& \word{running}\ |\ \word{final}\\
\nterm{kind} &::=& \{\word{mesh}\ |\ \word{contours}\ |\ \word{colors}\}\ 
     [ \word{limits}\ (\nterm{expr}, \nterm{expr}) ]\\
  &|& \word{quivers}\ [\nterm{primitive}\ \word{mesh}]\\
\nterm{options} &::=& \nterm{characters}\ \nl
\end{eqnarray*}
For example, the following command displays a scalar field $C$ as a heat map during the simulation:
\begin{lstlisting}
    display running C as colors
\end{lstlisting}
The following displays a vector field V as an array of quivers on a $0.2 \times 0.2$ mesh at the completion of the simulation:
\begin{lstlisting}
    display final V as quivers 0.2 mesh
\end{lstlisting}
A running display of a field can be converted to a movie, e.g.,
\begin{lstlisting}
    make movie Cvid.mp4 of C as contours
\end{lstlisting}
Finally, \Morphgen\  provides several commands for assessing the numerical stability of the simulation:
\begin{eqnarray*}
\nterm{stability report} &::=&  \word{report\ diffusion\ number\ for}\ \nterm{primitive}\\
  &|& \word{report\ Courant\ number\ for}\ \nterm{primitive}\\
  &|& \word{report\ Peclet\ number\ for}\ \nterm{primitive}\ \word{and}\ \nterm{primitive}
\end{eqnarray*}


\section{Example: Routing Neural Pathways}
Suppose we wanted to assemble a neuromorphic computer with a complexity comparable to a mammalian brain.
Such a neurocomputer might have billions of artificial neurons, each with many thousands of connections,
and it would be organized into functional regions, each comprising many millions of neurons, with remote regions connected by dense bundles of millions of neural fibers (artificial axons).
In a developing mammalian brain, axons grow toward their destinations by following chemical signals (morphogens) that indicate way stations and the final destinations.
The stepwise development of a morphogenetic program to solve this problem illustrates our approach to swarm intelligence.

\GenFig{swarm}{routing-5x5000-7.png}{width=\textwidth}{
Neural fiber bundle routing by modified flocking algorithm.
There are five bundles, each comprising 5000 fibers, joining randomly selected origins and destinations on the lower and upper surfaces.}

\subsection{Discrete Agents}
We begin with a simple path routing process.
An agent will be placed at the connection's origin (e.g., the originating point for a neural fiber), 
and a source of a diffusing morphogen (the attractant) will be placed at its destination.
If the agent moves up the attractant gradient, then it will find its way to the destination, and it can create in its wake the path from the origin to the destination.
However, we want to create many such paths, and so they cannot go in a straight line to their destinations, but must weave their way around already created paths.
There are at least two ways to accomplish this.
One is to have existing paths emit a repellant morphogen, which the agents try to avoid while seeking the attractant \Citepar{CMRS}.
An alternative, simpler solution is to have the existing paths absorb or degrade the morphogens, so that they become sinks for it \Citepar{MCHSA,CSMA}.
Although both solutions work, in this case the simpler solution works better, since the concentration of diffusing morphogens decreases exponentially with distance from the source \Citepar{MPPFCF}.
(For a 3-dimensional diffusion-decay process
$\dot{C} = D \del^2 C - C/\tau$,
the steady-state concentration at a distance $r$ from a point source with rate $k$ is
$ C(r) = \frac{k }{4 \pi D r} \exp(-r/\sqrt{D \tau})$.)
Since the concentration of attractant varies significantly with distance from the destination, it is difficult, with the first approach, to properly balance it against the repellant throughout the space.
This balance is achieved in the second approach because the existing paths absorb a fraction of the attractant in their vicinity.

Our goal, however, is not to have single neural fibers connecting (artificial) brain regions, but to have dense bundles of point-to-point connections.
Therefore, rather than having a single agent follow the attractant gradient from the origin to the destination, we want a large swarm of agents to do so, creating in its wake a bundle of fibers that is tight but not too tight.
Flocks of birds and schools of fish move in an organized way in compact groups, and so we have explored a modified flocking algorithm as a way of coordinating a swarm to lay down fiber bundles from an origin region to a destination region
\Citepar{CMRS,CSMA}.
This algorithm has been shown to scale up from five agents (and hence five fibers per bundle) up to 5000 agents and hence 5000 fibers per bundle, with no change of parameters and a small number of errors (acceptable in neural computation): scaling over four orders of magnitude \Citepar{CSMA}.
See Fig.\ \ref{fig:swarm}.

\subsection{Continuous Swarm}
Next, we take the number of agents in the swarm to the continuum limit.
The development and refinement of the morphogenetic algorithm is described in \textcite{MPPFCF};
here we summarize the final algorithm.

The path routing algorithm makes use of four substances: an attractant morphogen, the goal material marking the destination, the path material laid down by the agents, and the swarm substance, which is composed of agents.
The \emph{morphogen} substance is defined by a concentration field and a diffusion-decay equation, which describes its physical behavior:
\begin{lstlisting}[frame=single]
substance morphogen:
    scalar field A        // morphogen concentration
  behavior:
    param d_A = 0.03      // diffusion constant
    param tau_A = 100     // decay time constant
    D A += d_A * del^2 A - A/tau_A  // diffusion + decay
\end{lstlisting}
The change in $A$ is described by a partial change equation because this reflects only its physical diffusion and decay; it is also emitted by the \emph{goal} material and absorbed by the \emph{path} material.

The \emph{goal} material emits the attractant morphogen (and in this sense is active), but otherwise does not change:
\begin{lstlisting}[frame=single]
substance goal_material:
    scalar field G        // density of goal material
  behavior:
    param k_G = 100       // attractant release rate
    D G = 0               // G field is fixed
    D A += k_G*[G*(1-A)]  // goal emits attractant
\end{lstlisting}
The partial change equation for $A$ describes its emission by the \emph{goal} material up to saturation at $A=1$ and therefore acts as a source term.

    The \emph{path} material $P$, which is laid down by the swarm, has two properties:
(1) It absorbs the attractant morphogen (so that new paths avoid existing paths),
which is defined by a partial change equation for the morphogen,
$\Change A \decrby PA/\tau_P$.
Therefore the complete PDE for the morphogen is:
\[ \dot{A} = d_A  \del^2 A - A/\tau_A + k_G G(1-A) - PA/\tau_P. \]
(2) $P$ increases its concentration autocatalytically if its concentration is above a threshold $\theta_P$, and it decays if it is below the threshold, which is described by the partial change equation:
\neqn{autocat}{ \Change P \incrby a_P[P>\theta_P]P(1-P)-[P \leq \theta_P]P/t_P. }
This autocatalytic behavior sharpens up the paths and ensures they have consistent density, which the equation drives to $P=1$ (path present) or $P=0$ (path absent).
It is a partial equation because the swarm also deposits path material.
The foregoing behavioral specifications are combined in the \emph{path\_material} substance definition:
\PreChap{\newpage}{} 
\begin{lstlisting}[frame=single]
substance path_material:
    scalar field P       // path density
  behavior:
    params:
      a_P = 20       // autocatalytic rate
      theta_P = 0.3  // autocatalytic threshold
      theta_C = 0.02 // quorum threshold
      t_P = 1        // path decay time constant
      tau_P = 0.2    // attractant absorption time
    // autocatalysis:
    D P = a_P*[P>theta_P][P*(1-P)] - [P<=theta_P]P/t_P
    D A -= [P*A]/tau_P   // path absorbs attractant
\end{lstlisting}

Finally we define the \emph{swarm} substance, which controls the mass of agents to lay down a path while approaching the goal and avoiding existing paths.
The principal variable is the swarm density $C$, which is a real number reflecting the continuum approximation.
The \emph{swarm} substance also has a vector field $\Vvec$, which defines the velocity of the swarm throughout space.
The change in swarm density, then, is simply the negative divergence of the flux $C\Vvec$:
\[ \Change C = -[t>t_D] \divergence C\Vvec .\]
The conditional factor $[t>t_D]$ prevents swarm movement until $t_D$ time units have passed, which allows the morphogen to diffuse throughout the space.

The velocity field is a weighted combination of three influences: the normalized morphogen gradient, a density control term, and random (Brownian) motion to break symmetries.
The normalized morphogen gradient (or gradient versor) $\Vvec_1$ is
$\del A / \norm{\del A}$, but to avoid possible division by zero, we compute it:
$\Vvec_1 = \Uvec /(\norm{\Uvec}+\epsilon)$,
where 
$\Uvec = \del A$ and
$\epsilon$ is a small number.
To control the density, we define a potential function $(C-1)^2$ that is minimized at the desired density $C=1$.
The density control velocity is the negative normalized gradient of this potential:
$\Vvec_2 = -\Wvec/(\norm{\Wvec}+\epsilon)$,
where $\Wvec = \del[(C-1)^2]$.
A regularization parameter $\lambda$ controls the relative magnitude of these two velocity fields, 
$(1-\lambda)\Vvec_1 + \lambda \Vvec_2$, which are combined with Brownian motion $k_W \DW^2$ and an overall speed $v$ to compute the final velocity field:\footnote{
Alternately, the Brownian motion can be described by a diffusion term.
}
\neqn{reg}{ \Vvec = v [(1 - \lambda)\Vvec_1- \lambda \Vvec_2 + k_W DW^2)] .}
In \Morphgen\  the substance definition begins like this:
\begin{lstlisting}[frame=single]
substance swarm:
    scalar field C       // swarm density
    vector fields:
      U                  // morphogen gradient
      V                  // swarm velocity
      W                  // density gradient
  behavior:
    params:
      v = 1          // base swarm speed
      lambda = 0.1   // density regulation
      eps = 1e-100   // minimum gradient norm
      k_W = 0.3      // degree of random motion
      t_D = 5        // time delay
    let U = del A        // morphogen gradient
    let W = del[(C-1)^2] // density gradient
    let V = v * ((1 - lambda)*[U/(||U||+eps)] ...
        - lambda*[W/(||W||+eps)] + [k_W DW^2])
    D C = [t>t_D] -div[C*V] 
\end{lstlisting}
Finally and most importantly the swarm deposits path material at a relative rate $k_P$ but saturates at $P=1$.
Therefore the swarm contributes to the change in $P$ by 
$\Change P \incrby [t>t_D] k_P C(1-P)$, which begins after $t_D$ time units.
This is expressed in \Morphgen :
\begin{lstlisting}
    param k_P = 30       // path deposition rate
    D P += [t>t_D] k_P*[C*(1-P)] // path deposition
\end{lstlisting}

To test the algorithm we can define some obstacles and an origin and destination for a new path.
Everything we have discussed so far works for either a two- or three-dimensional morphogenetic system, but there are minor differences between a 2D and 3D simulation.

For a 2D simulation, the obstacles (representing previously generated paths) are just circular regions of path material.
These can be created by a $\word{body}$ definition such as shown in Section~\ref{sec:Bodies}.
For 3D simulations there are additional ways to create obstacles, or the algorithm itself can be used to generate multiple paths, as in the earlier simulations with discrete swarms.

To initialize the creation of a path, the swarm needs to be placed at the origin, and goal material needs to be placed at the destination.
Here are typical body definitions:
\begin{lstlisting}[frame=single]
body Cohort of swarm:
  for -0.05 < x < 0.05, -0.95 < y < -0.9: C = 1

body Goal of goal_material:
  for -0.05 < x < 0.05, 0.9 < y < 0.95: G = 1
\end{lstlisting}

A series of simulations exploring algorithm variants and parameter values is described in \textcite{MPPFCF}; here we present typical results in Figure \ref{fig:paths}.

\MultiFig{paths}{
\SubFig{2Dpath}{180608-2_18-52_.pdf}{width=0.48\textwidth}%
{2D simulation of path formation, origin at lower right, destination at upper left, 9 obstacles.}
\hfill
\SubFig{3Dpath}{P3Dt7.pdf}{width=0.48\textwidth}%
{3D simulation of path formation, origin in foreground, destination on back surface, four obstacles.
The figure shows regions of path density $P>0.5$.}
}{Simulations of neural path formation.
In both simulations it can be seen that a swarm sometimes splits to go around an obstacle.
For more information, see \textcite{MPPFCF}.
}

Figure \ref{fig:2Dpath} shows the result of a 2D simulation in which the origin was placed near the lower right corner and the destination was placed near the upper left.
The nine disks represent previous paths, which are obstacles to be avoided.
The new path finds its way around the obstacles, splitting to do so (similar to the bundles in Fig.\ \ref{fig:swarm}).
The degree of allowed splitting can be controlled by the swarm density regularization parameter $\lambda$ (Eq.\ \ref{eq:reg}).
The edges of the bundle are sharp and the density is constant due to the autocatalysis of path material
(Eq.\ \ref{eq:autocat}).

Figure \ref{fig:3Dpath} shows the result of a 3D simulation in which the origin was placed near the foreground and the destination was placed on the rear surface.
The path found its way around four obstacles representing previously generated paths.
The 2D and 3D morphogenetic programs are essentially identical.
     
\GenFig{body+legs-SAPalpha}{SAPalpha.png}{width=\textwidth}{
Simulation of swarm assembly of insect-like robot body.
The body is growing toward the right; the head segment is visible on the left, and the tailbud is at the extreme right but not shown.
Between them, eight segments have been assembled, each with a pair of segmented legs (the last two pairs incomplete at this point in the simulation, $T=60$).
}
\section{Example: Body and Leg Segmentation}\label{sec:segmentation}
Our second example illustrates how a swarm of microscopic agents can be coordinated to produce a simple insect-like robot body with a segmented ``spine'' and segmented ``legs.''
(A typical 2D result is shown in Fig.\ \ref{fig:body+legs-SAPalpha}.)
The algorithm uses a process inspired directly by embryological spinal development but it uses it in two different ways: to segment the robot's spine (similar to its function in vertebrate development) and to segment the robot's legs (which is not how insects' or other animals' legs are segmented).
Thus it shows how natural morphogenetic processes can be redeployed for different purposes in artificial morphogenesis.

The process in question is the \emph{clock-and-wavefront model} of spinal somatogenesis, which was first proposed in 1976 and finally confirmed in 2008
\Citepar{Cooke-Zeeman,Dequeant-Pourquie}.
As the vertebrate embryo develops, a pacemaker in its tailbud periodically produces a pulse of chemical, a segmentation morphogen.
This chemical pulse propagates toward the head of the embryo, transmitted through the tissue, as each cell is stimulated by its neighbors to produce its own pulse.
As each pulse passes through undifferentiated tissue, it causes the differentiation of one more somite or spinal segment.
This takes place in a sensitive region defined by a low concentration of two morphogens: a \emph{caudal morphogen}, which diffuses towards the head from the tailbud, and a \emph{rostral morphogen}, which diffuses toward the tail from already differentiated segments.
Therefore, the segments differentiate one by one from the head end toward the tail as the embryo grows.
The ratio of the growth rate to the pacemaker frequency determines the length of the segments, while the product of the frequency and growth duration determines the number of segments (which is characteristic of a species).

\subsection{Growth Process}\label{sec:growth process}
Our morphogenetic program to assemble the robot body will make use of three different substances.
The most important one is called \emph{medium} and represents the mass of active components (the swarm) that does most of the work.
This substance has a variety of properties, represented by scalar and vector fields;
the most important is its density $\Medium$;
others will be introduced as required below.
When it is part of a defined segment, \emph{medium} is in a differentiated state, and the density of differentiated tissue is represented by a scalar field $S>0$.

As in biology, differentiation means a change of internal state variables that affects the behavior and other properties of a substance.
In this case, the differentiation into segment tissue could cause the agents to couple with their neighbors, resulting in a rigid structure, or to secrete some structural substance.

The second substance represents \emph{terminal} tissue: either the tailbud of the developing spine or the ``feet'' of the developing legs;
its principal property is a scalar field $\Tail$, its density. 
Terminal tissue is composed of active components, and it is responsible both for growth processes and for generating the pacemaker signals.
During the morphogenetic process, terminal tissue has a velocity: away from the head during spinal development, and away from the spinal axis during leg development.

The initial state of the morphogenetic process requires the preparation of two bodies:
a head segment already in a differentiated state, and a tail segment adjacent to it.
For a 2D simulation, this could be written:
\begin{lstlisting}[frame=single]
body Head of medium:
  for 0.01 < x < 1, -0.5 < y < 0.5:
    M = 1    // medium tissue
    S = 1    // segment tissue
    // other initialization

body Tail of terminal:
  for 1 < x < 2, -0.75 < y < 0.75:
    T = 1    // initial tail bud
    // other initialization
\end{lstlisting}

As the terminal tissue moves away from medium tissue, the intervening space is filled with new, undifferentiated medium tissue.
In biological development this is a result of cell proliferation, which expands the mass of undifferentiated tissue and pushes the tailbud further away from the head.
In previous versions of our morphogenetic program, we have given the tail an initial velocity away from the head segment, and as it has moved it has produced undifferentiated tissue in its wake
\Citepar{MMNC,EC-APCAM,MCHSA,CMRS,CSMA}.
This approach makes sense if the agents are able to reproduce themselves, for example, if they are implemented by genetically engineered microorganisms.
If the agents are microrobots, however, and presumably unable to reproduce, then we need an alternative approach to growth.
In this case the agents can be supplied from an external source and routed to the growth region.
As tail tissue moves, it emits an attractant morphogen into the growth area (the growing gap between the existing body and the tail).
Available agents from the external source follow the attractant gradient into the growth area until they reach a desired density ($\Medium=1$ in our case).
Figure \ref{fig:growth} shows a simulation of growth in progress.

\def\thirdwidth{0.33}
\MultiFig{growth}{
\SubFig{growthT}{growthS2_T.png}{{width=\thirdwidth\textwidth}}{Terminal tissue $\Tail$.}
\hfill
\SubFig{growthH}{growthS2_H.png}{{width=\thirdwidth\textwidth}}{Attractant $\GrowthAttractant$.}
\hfill
\SubFig{growthM}{growthS2_M.png}{{width=\thirdwidth\textwidth}}{Undifferentiated tissue $\Medium$.}
}{Simulation of spine growth process in progress ($t=30$).
(a) Density $\Tail$ of terminal tissue (moving to right).
(b) Concentration $\GrowthAttractant$ of attractant emitted by terminal tissue.
(c) Density $\Medium$ of assembled undifferentiated tissue.
Agents are introduced by the sources along the upper and lower edges and migrate up the attractant gradient to assemble into the spine.
}

This approach requires a source of free agents (particles of \emph{medium}), and so we define a substance \emph{source} to represent this process; it has a scalar field $\Source$ representing the density of source material ($\Source \in \{0, 1\}$).
The source provides agents at rate $\Const{\Source}$ up to saturation ($\Medium=1$);
therefore, as agents move out of the source regions, more will be introduced to maintain $M=1$ density at the source.
This provision of free agents takes place only so long as growth continues, which is represented by the condition $\GrowthTimer > \Growththresh$.
Therefore the source of new agents is described
$\Change\Medium \incrby [\GrowthTimer > \Growththresh] \Const{\Source} \Source(1-\Medium)$.
The \emph{source} substance is defined like this in \Morphgen :
\begin{lstlisting}[frame=single]
substance source:
    scalar field J    // source density
  behavior:
    param k_J = 1     // input rate
    D M += [G > theta_G] k_J * [J * (1 - M)]
\end{lstlisting}

Source regions must be prepared as part of the initialization of the morphogenetic process.
For example, the simulation in Figure \ref{fig:growth} uses the following sources:
\begin{lstlisting}[frame=single]
body AgentSources of source:
  for 5 < x < 6, 1.8 < y < 1.95: J = 1   // upper source
  for 5 < x < 6, -1.95 < y < -1.8: J = 1 // lower source
\end{lstlisting}

Terminal tissue has the following behavior.
Let $\TailDirection$ be a unit vector field pointing in the direction of desired growth,
and let $\TailRate$ be the desired growth rate, which is the same as the speed of terminal tissue movement.
Then the change in terminal concentration is the negative divergence of its flux:
\[
 \Change \Tail = - \divergence (\Tail \TailDirection [\GrowthTimer > \Growththresh] \TailRate) .
\]
The condition $\GrowthTimer > \Growththresh$ holds so long as growth continues; when it fails, the movement rate is effectively zero and the terminal tissue stops moving.
Figure \ref{fig:growthT} shows the massive swarm of terminal agents moving to the right.

Terminal tissue emits a growth attractant, with concentration $\GrowthAttractant$, up to saturation,
which diffuses and decays at specified rates:
\begin{eqnarray*}
 \Change \GrowthAttractant &\incrby& \Const{\GrowthAttractant} \Tail (1 - \GrowthAttractant) ,\\
 \Change \GrowthAttractant &\incrby& \DiffRate{\GrowthAttractant} \del^2 \GrowthAttractant
   - \GrowthAttractant / \DecayRate{\GrowthAttractant} .
\end{eqnarray*}
Figure \ref{fig:growthH} shows the diffusion of the attractant from the terminal tissue.

The swarm of agents $\Medium$ from the sources follows the attractant gradient $\del\GrowthAttractant$ at a speed controlled by $\GrowthRate$ and aggregating up to a maximum density of $\Medium=1$:
\neqn{DM}{
 \Change\Medium = -\GrowthRate  [\Medium < 1] \divergence [\Medium \del \GrowthAttractant] .
}
Once the agents have assembled in the spine (where $\Medium=1$) they do not move again.
Figure \ref{fig:growthM} shows the growth of undifferentiated spinal tissue in progress.\footnote{
Other simulation results in this chapter use the original growth process.
}

\GenFig{SAPalpha-progress}{SAPalpha-progress.png}{width=\textwidth}{
Diffusion of $\Posterior$ and $\Anterior$ morphogens after seven new spinal segments have differentiated (tan color).
Blue color at  far right represents $\Posterior$ (caudal morphogen) diffusing from the tailbud.
Green color around the segments represents $\Anterior$ (rostral morphogen) diffusing from differentiated segments.
The $\Smorphogen$ wave visible between the sixth and seventh segments has just differentiated the seventh segment (which is not yet emitting $\Anterior$).
The developing legs are also emitting $\Posterior$ and $\Anterior$ morphogens.
}

\subsection{Spine Assembly}
The foregoing explains the growth of undifferentiated spinal tissue; 
we turn now to the clock-and-wavefront process, which causes the spine to differentiate into segments of specified number and size.

The concentrations of the caudal and rostral morphogens are represented by scalar fields $\Posterior$ and $\Anterior$, respectively
(Fig.\ \ref{fig:SAPalpha-progress}).
The rostral morphogen is produced by already differentiated segment tissue (represented by segment density $\Somites>0$), from which it diffuses and decays:
\neqn{DA}{
 \Change \Anterior =
  \AntSat \Somites (1 - \Anterior)
  + \AntDiff \Lapl \Anterior - \Anterior / \AntDecay .
}
The caudal morphogen concentration $\Posterior$ is considered a property of the \emph{medium} substance, and its diffusion and decay are physical properties defined by a partial equation in the behavior of \emph{medium}:
\[
 \Change \Posterior \incrby
  \PstDiff \Lapl \Posterior - \Posterior / \PstDecay .
\]
The caudal morphogen is produced from terminal tissue (i.e., tissue with terminal density $\Tail>0$), which is part of the behavior of the  \emph{terminal} substance:
\[
 \Change \Posterior \incrby
  \PstSat \Tail (1 - \Posterior)  .
\]

\GenFig{alpha}{alpha.png}{width=\textwidth}{
Propagation of $\Smorphogen$ morphogen.
Two successive waves are propagating leftward toward the head.
The decaying $\Smorphogen$ pulse in the tailbud is visible at the righthand end.
}

The segmentation signal $\Smorphogen$ is produced periodically in the tailbud and is actively propagated forward through the \emph{medium} tissue (Fig.\ \ref{fig:alpha}).
The segmentation morphogen is a property of \emph{medium}, which defines its physical diffusion and decay rates:
\[
 \Change\Smorphogen \incrby \DiffRate{\Smorphogen} \del^2 \Smorphogen
  - \Smorphogen / \DecayRate{\Smorphogen} .
\]
Terminal tissue ($\Tail>0$) in the tailbud produces a pulse of $\Smorphogen$ so long as growth is continuing (represented by $\GrowthTimer > \Growththresh$) and the tail's pacemakers are in the correct phase (represented by $\Clock > \Clockthresh$):
\[
 \Change\Smorphogen \incrby \Const{\ClockCond}
   [\GrowthTimer > \Growththresh \wedge \Clock > \Clockthresh]
   \Tail(1-\Smorphogen) .
\]
This partial equation is an extension of $\Smorphogen$ behavior in the definition of the \emph{terminal} substance.
(Additional detail on the pacemaker and growth duration can be found in 
\textcite{CSMA} and \Secref{leg assembly} below.)

The \emph{medium} substance also produces $\Smorphogen$ when it is sufficiently stimulated by it ($\Smorphogen > \Sthresh$), which allows it to propagate a wave of $\Smorphogen$ stimulation.
After it produces an $\Smorphogen$ pulse, the tissue enters a refractory period (represented by $\Rec \geq \Recthresh$), which ensures that the wave is unidirectional (towards the head).
The variable $\FireCond$ (which is only transiently nonzero) represents the density of \emph{medium} tissue in this sensitive and stimulated state, and is used both to produce a pulse of $\Smorphogen$ and to start the refractory timer:
\begin{eqnarray}
 \label{eq:FireCond}
 \FireCond &=& [\Smorphogen > \Sthresh \wedge \Rec < \Recthresh] \Medium ,\\
 \Change\Smorphogen &\incrby& \Const{\FireCond}\FireCond(1-\Smorphogen) ,\\
 \Change\Rec &=& \FireCond - \Rec / \RecDecay .
\end{eqnarray}
         
The actual differentiation of a region of \emph{medium} tissue into a segment ($\Somites>0$) takes place when a sufficiently strong ($ \Smorphogen > \SLwb$) segmentation wave passes through a region with sufficiently low rostral ($\Anterior < \AntUpb$) and caudal ($\Posterior < \PstUpb$) morphogen concentrations.
This causes a rapid $\Constant{\SomiteTrigger} (1 - \Somites)$ increase in differentiated tissue, which then triggers autocatalytic differentiation
$\SomiteRate\Somites(1-\Somites)$ until it is complete ($\Somites=1$):
\begin{eqnarray*}
 \Change \Somites &=&
        \SomiteRate
       \Somites  (1 - \Somites)
       +
       \If{ \Smorphogen > \SLwb \wedge \Anterior < \AntUpb \wedge \Posterior < \PstUpb}
       \Constant{\SomiteTrigger} (1 - \Somites) \\
      &=&  (\SomiteRate
       \Somites 
       +
       \If{ \Smorphogen > \SLwb \wedge \Anterior < \AntUpb \wedge \Posterior < \PstUpb}
       \Constant{\SomiteTrigger} ) (1 - \Somites) .
\end{eqnarray*}
This can be written as follows in the current version of the \Morphgen\  language:
\begin{lstlisting}
  let chi = [alpha > alpha_lwb][R < R_upb][C < C_upb]c_seg
  D S = [(chi + k_S * S) * (1 - S)]
\end{lstlisting}

\GenFig{seg3D-S}{seg3D-S.png}{width=\textwidth}{
Three-dimensional simulation of segmentation.
Three new segments posterior to the head have been assembled.
The figure shows regions in which $\Somites > 0.5$.
}

There is a small gap between the differentiated segments because segmentation takes place in a region where the rostral concentration $\Anterior$ has dropped to $\AntUpb$
(see Figs.\ \ref{fig:body+legs-SAPalpha}, \ref{fig:SAPalpha-progress}).
To determine the width $\SegGap$ of the gap, set the threshold  $\AntUpb$ equal to the concentration at a distance $\SegGap$ from the anterior differentiated tissue: in a 2D space,
$\AntUpb = \frac{\AntSat}{ 2 \pi \AntDiff} K_0(\SegGap / \sqrt{\AntDiff \AntDecay}) $,
where $K_0$ is a modified Bessel function of the second kind (Basset function).
Figure \ref{fig:seg3D-S} shows a 3D simulation of segmentation using the same morphogenetic program and parameters as the 2D simulation.
       
\GenFig{BAM-BPM-alpha}{BAM-BPM-alpha.png}{width=\textwidth}{
Diffusion of $\AntBordMorph$ and $\PostBordMorph$ morphogens from anterior and posterior tissue, respectively, in each differentiated segment.
Orange color represents $\AntBordMorph$ morphogen diffusing from differentiated anterior tissue,
and green color represents $\PostBordMorph$ morphogen diffusing from differentiated posterior tissue.
White lines on the spinal surface are imaginal tissue $\ImagDisk$ from which the legs grow.
(Two $\Smorphogen$ waves are also visible.)
}

\subsection{Leg Assembly}\label{sec:leg assembly}
The process described so far is analogous to the somitogenesis or spinal segmentation that takes place during vertebrate embryological development.
Now however we exploit the process for a different morphogenetic task: the generation of a pair of segmented legs on each spinal segment (e.g., Fig.\ \ref{fig:body+legs-SAPalpha}).

The first task is to control the placement of the legs on the spinal segments;
in the case of a 2D simulation, the only variable is their location along the length of a segment.
This can be determined by the relative concentration of morphogens that diffuse from the anterior and posterior ends of each segment (Fig.\ \ref{fig:BAM-BPM-alpha}), but this requires distinguishing the anterior and posterior ends of each segment: a symmetry breaking process.

The required information is available when the segmentation $\Smorphogen$ wave passes through the tissue, since at that time the more anterior tissue is emitting the rostral morphogen $\Anterior$, and the more posterior tissue is emitting the caudal morphogen $\Posterior$.
Therefore, the anterior region of a segment further differentiates into anterior tissue (with density $\AntBorder$) when the segmentation wave passes through ($\Smorphogen > \LWB{\Smorphogen}$) and the rostral morphogen is in the correct range ($0.5 \AntUpb > \Anterior > 0.25 \AntUpb$ in this simulation),
which trigger an autocatalytic differentiation process that goes to completion:
\[ \Change \AntBorder
 = \Const{\AntBorder} \Somites \AntBorder (1 - \AntBorder)
 + \If{ 0.5 \AntUpb > \Anterior > 0.25 \AntUpb
    \wedge \Smorphogen > \LWB{\Smorphogen}} \Constant{\AntBorder}
 - \AntBorder / \DecayRate{\AntBorder} .
\]
Here, we expect $\Anterior$ to reflect rostral morphogen diffusing from segments anterior to the currently differentiating segment.
However, the differentiation of $\AntBorder$ tissue takes place simultaneously with the differentiation of $\Somites$ tissue (as $\Smorphogen$ passes through), which begins immediately to emit $\Anterior$ morphogen.
This will cause the entire segment to differentiate into $\AntBorder$ tissue,
and to prevent this we introduce a state variable $\AntBlock$ that temporarily blocks production of the rostral morphogen until anterior tissue differentiation is complete.
It is set by the the segmentation impulse (Eq.\ \ref{eq:FireCond}) and then decays exponentially:
\[
 \Change \AntBlock =
  \Constant{\AntBlock} \FireCond - \AntBlock / \DecayRate{\AntBlock}
   + [\AntBlock > 1](1-\AntBlock) / t_\AntBlock .
\]
We choose the time constant $\DecayRate{\AntBlock}$ to allow $\AntBorder$ differentiation to complete.
We replace Eq.\ \ref{eq:DA} for the production of $\Anterior$ with:
\[
 \Change \Anterior =
  \AntSat [\AntBlock < \theta_\AntBlock] \Somites (1 - \Anterior)
  + \AntDiff \Lapl \Anterior - \Anterior / \AntDecay ,
\]
which blocks production until $\AntBlock$ has decayed below $\theta_\AntBlock$.

Similarly, the $\Smorphogen$ wave and caudal morphogen in an appropriate range ($0.95 \PstUpb > \Posterior > 0.8 \PstUpb$) trigger autocatalytic differentiation of posterior border tissue:
\[ \Change \PostBorder
 = \Const{\PostBorder} \Somites \PostBorder (1 - \PostBorder)
 + \If{ 0.95 \PstUpb > \Posterior > 0.8 \PstUpb
    \wedge \Smorphogen > \LWB{\Smorphogen} } \Constant{\PostBorder}
 - \PostBorder / \DecayRate{\PostBorder} .
\]
(Emission of caudal morphogen does not have to be blocked since it comes from the tailbud.)

Once the anterior and posterior border tissues have differentiated, they can do their job of emitting anterior and posterior border morphogens, which provide a reference frame for determining position along each segment:
\begin{eqnarray*}
\Change \AntBordMorph
 &=& \If{\AntBorder > \Threshold{\AntBorder}}\Const{\AntBordMorph} \Somites 
   (1 - \AntBordMorph)
  + \DiffRate{\AntBordMorph} \Lapl \AntBordMorph
  - \AntBordMorph / \DecayRate{\AntBordMorph} ,
  \label{eq-Da}
\\
\Change \PostBordMorph
 &=& \If{ \PostBorder > \Threshold{\PostBorder}}\Const{\PostBordMorph} \Somites 
   (1 - \PostBordMorph)
  + \DiffRate{\PostBordMorph} \Lapl \PostBordMorph
  - \PostBordMorph / \DecayRate{\PostBordMorph} .
  \label{eq-Dp}
\end{eqnarray*}
The thresholds ($\AntBorder > \Threshold{\AntBorder}$, $\PostBorder > \Threshold{\PostBorder}$) ensure that only well-differentiated tissue produces these morphogens.
         
Our plan is for the legs to be assembled on \emph{imaginal tissue} at the correct location on each segment.
The anterior/posterior position is defined by anterior and posterior border morphogens in the correct ranges 
($\UPB{\AntBordMorph} > \AntBordMorph >  \LWB{\AntBordMorph}$,
$\UPB{\PostBordMorph} > \PostBordMorph > \LWB{\PostBordMorph}$),
but these morphogens diffuse throughout the spinal tissue and we want to ensure that imaginal tissue differentiates only on the surface of the spine.
Moreover, this notion of being ``on the surface'' should be independent of the agent size (which is in fact infinitesimal in our continuum model), so that the morphogenetic algorithm is scale-invariant.
Taking a cue from the natural world, in which bacteria and other organisms do \emph{quorum sensing} to determine if their population density is sufficient for some purpose, we have our agents estimate their local population density.
If it is near its maximum ($M \approx 1$ in our case), then an agent knows it is in the interior;
if it is near its minimum ($M \approx 0$) then the agent is relatively isolated.
If however the local density is $M \approx 0.5$, then the agent knows it is near the surface;
how large $|M-0.5|$ is allowed to be will determine the effective thickness of the surface layer.

This strategy requires that an agent be able to determine its local population density.
There are many ways to do this, but perhaps the most practical is for the agents to emit a slowly diffusing, rapidly decaying morphogen $\SmoothDensity$, which serves to broadcast density information over a short range.
The local morphogen concentration becomes a surrogate for population density, more precisely, for the density convolved with a smoothing kernel.
(See \Secref{Morphgen-compiler} for more on implementation of this strategy.)
Morphgen code such as the following can be used to define a scalar field $\Edge$ that represents closeness to the surface (or in the case of 2D simulations, the edge).
\begin{lstlisting}
  D N = k_N*[M*(1-N)] + D_N*del^2 N - N/tau_N // density surrogate
  let E = [N > 0.15] [N < 0.2] S              // edge marker
\end{lstlisting}
Then the autocatalytic differentiation of imaginal tissue can be triggered by a combination of correct morphogen ranges and being near to the surface:
\neqn{imaginal}{
 \Change \ImagDisk
 = \If{ \UPB{\AntBordMorph} > \AntBordMorph >  \LWB{\AntBordMorph}
  \wedge \UPB{\PostBordMorph} > \PostBordMorph > \LWB{\PostBordMorph} }
   \Edge \Somites (1 - \ImagDisk) .
}
This equation is written as follows in \Morphgen :
\begin{lstlisting}
  D I = [a_upb > a][a > a_lwb][p_upb > p][p > p_lwb] [E*[S*(1-I)]]
\end{lstlisting}

Once the imaginal tissue has fully differentiated, the next step is to initialize the clock-and-wavefront process to grow and segment the legs.
This is accomplished by having the imaginal tissue differentiate into terminal tissue with a velocity vector directed outward from the spine; thus the imaginal tissue becomes the ``foot'' of a future leg.
We have used the rapid differentiation of imaginal tissue ($\Change\ImagDisk > \IDchangethresh$) to trigger this transformation.
The velocity vector is directed down the $S$ gradient: $-\del \Somites / \norm{\del\Somites}$.
The leg grows just like the spine grows, recruiting free agents from the sources to fill in the gap between the developing leg and the outward moving foot.
The exact same morphogens
($\GrowthAttractant$, $\Anterior$, $\Posterior$, $\Smorphogen$, $\AntBordMorph$, $\PostBordMorph$)
and tissue types
($\Tail$, $\Somites$, $\AntBorder$, $\PostBorder$)
are used to control leg segmentation as to control spinal segmentation.
         
\subsection{Termination}
There are two problems with using the same morphogenetic process for the spine and for the legs.
The first is that with the same parameters, the number and size of the segments will be the same in the spine and the legs, which might not be what we want.
The second is that, in the absence of a mechanism to prevent it, the leg segments will develop their own imaginal tissue, from which ``leglets'' will grow, and so on, in a fractal manner, which is not our goal.
Therefore we must consider the termination of morphogenetic processes, which in fact is an interesting problem in embryology.
How big should limbs and organs get? What limits growth processes in normal development?

\GenFig{omega2}{omega2.png}{width=\textwidth}{
Segmentation with higher pacemaker frequency.
This simulation has twice the pacemaker frequency $\ClockfreqHz_{\rm S}$ as the simulation shown in Fig.\ \ref{fig:SAPalpha-progress}.
Both are shown at the same simulation time, but this has assembled 12 segments (compared to six in the previous simulation) with ten pairs of legs compared to five previously;
the segments are also approximately half the length as in the previous simulation.
In this simulation $\ClockfreqHz_{\rm S} = 1/\pi$, the previous had $\ClockfreqHz_{\rm S} = 1/(2\pi)$.
}

In our morphogenetic program we have adopted a simple mechanism.
In several of the equations above, we have seen that a process continues only so long as a condition $\GrowthTimer > \Growththresh$ is satisfied.
Here $\GrowthTimer$ is a scalar property of terminal tissue, which decays exponentially:
$\Change\GrowthTimer = -\GrowthTimer / \GrowthDecay$,
where $\GrowthDecay$ is the growth time constant.
If the initial value of $\GrowthTimer$ is $\GrowthTimer_0$, then the $\Growththresh$  threshold will be reached at time $t =  \GrowthDecay \ln (\GrowthTimer_0 / \Growththresh)$.
If $\ClockfreqHz$ is the pacemaker frequency,
then $\lfloor \ClockfreqHz \GrowthDecay \ln (\GrowthTimer_0 / \Growththresh) \rfloor$ complete segments will be generated.
The length of the segments will be $\GrowthRate / \ClockfreqHz$, where $\GrowthRate$ is the growth rate (the speed of terminal tissue movement).
There are a number of parameters here that could be controlled, but if we want the same processes operating in spine and leg segmentation, then probably the simplest are the initial timer value and pacemaker frequency.
For spinal segments of length $\SpineSegLength$, we set the pacemaker frequency to
$\ClockfreqHz_{\rm S} = \GrowthRate / \SpineSegLength$.
For $\SpineSegNum$ spinal segments, we initialize the timer to
\[
 \GrowthTimer_{\rm S} 
  = \Growththresh \exp \left( \frac{\SpineSegNum}{\ClockfreqHz_{\rm S} \GrowthDecay} \right).
\]
See Fig.\ \ref{fig:omega2} for the effect of $\ClockfreqHz_{\rm S}$ on spinal segment length and number.
The equations for $\LegSegNum$ leg segments of length $\LegSegLength$ are analogous.

As mentioned above, rapid differentiation of imaginal tissue can be used as a transient impulse to initialize leg growth; we define it $\LegInit = [ \Change\ImagDisk > \IDchangethresh ] \Constant{L}$. This impulse can be used to trigger 
conversion of imaginal tissue to terminal tissue
and to initialize the leg timers and pacemaker frequencies:
\begin{eqnarray}
\label{eq:DLTerm}
\Change\Tail &\incrby& \Const{{\rm L}} \ImagDisk \Tail(1-\Tail) + \LegInit,\\
\Change \GrowthTimer &\incrby&
  \Const{{\rm GL}} \ImagDisk \LegInit (\GrowthTimer_{\rm L} - \GrowthTimer) ,\\
\Change \ClockfreqHz &\incrby&
  \Const{\ClockfreqHz} \ImagDisk \LegInit (\ClockfreqHz_{\rm L} - \ClockfreqHz) .
\end{eqnarray}
$\Const{{\rm GL}}$ and $\Const{\ClockfreqHz}$ are large so they reset the parameters while $\LegInit>0$.
Agents in imaginal tissue also reorient their velocity vectors down the $\Somites$ gradient, that is, outward from the spine:
\[
 \Change \TailDirection \incrby
  \ImagDisk\Const{{\rm u}} (-\del \Somites / \norm{\del\Somites} - \TailDirection).
\]

There are several ways to prevent the formation of fractal ``leglets'' on the legs; the most direct is to block the differentiation of imaginal tissue, as given by Eq.\ \ref{eq:imaginal}.
The problem is that this equation is part of the behavior of \emph{medium} tissue, which is the same in the spine and legs, since its constituent agents come from a common source.
Leg terminal tissue, however, develops from imaginal tissue, but tail terminal tissue does not, which provides a basis for distinguishing the legs from the spine.
Therefore we introduce a variable $\LegTissue$ representing the density of tissue in the leg state.
When the differentiation of imaginal tissue triggers the differentiation of leg terminal tissue (Eq.\ \ref{eq:DLTerm}), this tissue can simultaneously differentiate into the $\LegTissue$ state, which differentiates it from tail terminal tissue:
$\Change\LegTissue \incrby \Const{{\rm L}} \ImagDisk \LegTissue(1-\LegTissue) + \LegInit$.
As \emph{medium} particles arrive from the sources and assemble themselves between the foot terminal tissue and the growing legs, they must also inherit the $\LegTissue$ property.
This can be accomplished by having $\LegTissue$ tissue emit a short-range morphogen $\LegMorph$ that causes developing \emph{medium} tissue to inherit the $\LegTissue$ property.
Therefore, if the concentration of $\LegMorph$ in \emph{medium} tissue is above a threshold (meaning that it is near ``foot'' terminal tissue) and it is not spinal tissue, then it enters the $\LegTissue$ state:
\[
 \Change\LegTissue \incrby \Const{\LegInit} [\LegMorph > \Threshold{\LegMorph} \wedge \Somites < 0.5]  (1-\LegTissue).
\]
Then we change imaginal tissue differentiation (Eq.\ \ref{eq:imaginal}) so that it takes place only in non-leg tissue ($\LegTissue \approx 0$):
\[
 \Change \ImagDisk
 = \If{ \UPB{\AntBordMorph} > \AntBordMorph >  \LWB{\AntBordMorph}
  \wedge \UPB{\PostBordMorph} > \PostBordMorph > \LWB{\PostBordMorph} 
  \wedge \LegTissue < \Threshold{\LegTissue} }
   \Edge \Somites (1 - \ImagDisk) .
\]

Finally, note that the spinal tissue between the last segment and the tailbud is undifferentiated because the caudal morphogen concentration $\Posterior$ is too high.
Therefore it does not develop legs and forms the tail proper between the leg-bearing segments and the tailbud.
To determine the tail's length $\TailLength$, set the caudal morphogen concentration at a distance $\TailLength$ in front of the tailbud to the caudal morphogen threshold for segmentation: in 2D,
$\PstUpb = \frac{\PstSat }{2 \pi \PstDiff} K_0(\TailLength / \sqrt{ \PstDiff \PstDecay })$.
Also, as can be seen in 
Figs.\ \ref{fig:body+legs-SAPalpha}, \ref{fig:seg3D-S}, and \ref{fig:omega2}, 
the first spinal segment posterior to the head can have a different length compared to the other segments; this is because its length depends on the initial phase of the pacemaker.
     
\section{Toward a \MORPHGEN\ Compiler}\label{sec:Morphgen-compiler}

\subsection{Agents as Substantial Particles} 
Our approach to morphogenetic engineering treats agent swarms and other substances from the perspective of continuum mechanics.
That is, although our swarms and tissues are composed of finite numbers of discrete elements of finite size, our intended application is to very large numbers of very small elements, and so the continuum approximation is useful.
From the continuum mechanics perspective, tissues and massive swarms are infinitely divisible into infinitesimal \emph{material points} or \emph{particles} representing differential volume elements of continua \Citepar{CSMA}.
Therefore, a material point or particle does not represent a single agent (or molecule or other active or passive component), but a very tiny volume containing many such physical elements.
As a consequence, thinking about the individual behavior of particles in such a continuum is helpful in understanding how the individual agents should behave, but it is not sufficient to equate particles and agents.
For example, a particle might have a distribution of orientations or velocities, or be in a mixture of differentiation states, whereas individual agents have definite orientations, velocities, and differentiation states \Citepar{PDFECAM,MMAM,AMEEC}.

We have seen how PDEs can be used to describe the motion, differentiation, and other behavior of swarms of microscopic agents to assemble complex shapes, but these PDEs cannot be used directly to control the agents.
These PDEs have been expressed in an \emph{Eulerian} (spatial) reference frame, in which the derivatives are relative to fixed spatial locations and describe the changes of variables at those locations as particles flow through them.
For example, we may be concerned with the change in temperature $\aPropS(t,\anElemPos)$ at a particular location $\anElemPos$, that is, 
$\partial \aPropS(t,\anElemPos) / \partial t$, as particles with different temperatures flow through it.
More relevant for the control of agent swarms is a \emph{Lagrangian} (material) reference frame, in which derivatives are relative to fixed particles and describe how the properties of those particles change as they move through space.
Thus we might be more concerned with the temperature $\aPropM(t,\anElem)$ of a particle $\anElem$ and how it changes as that particle flows through space, $\partial \aPropM(t,\anElem) / \partial t$;
the latter is called the \emph{material} or \emph{substantial} derivative and is commonly written
$\D\aPropM / \D t$.

Since a particle's position is a function of its velocity $\anElemVel$, the material derivative can be expanded by the chain rule:
\[ \frac{\D \aPropM}{\D t} = \frac{\partial \aPropS}{\partial t} + \anElemVel \cdot \del \aPropS .\]
For a vector property of particles, the formula is analogous:
\[ \frac{\D \vect{\aPropM}}{\D t}
 = \frac{\partial \vect{\aPropS}}{\partial t} + \anElemVel \cdot \del \vect{\aPropS} ,\]
where $\del \vect{\aPropS}$ is a second-order tensor.
(In a Cartesian coordinate frame $(\del \vect{\aPropS})_{jk} = \partial \aPropS_j / \partial x_k$
and $\anElemVel \cdot \del \vect{\aPropS} = \anElemVel\trans \del \vect{\aPropS}$.)

\subsection{Morphogen-based SPH Control} \label{sec:toward}

Morphgen describes continuous fields of infinite numbers of infinitesimal agents.
A compiler for the Morphgen language would produce code for given numbers of agents of given finite size, and would thus realize the scalability potential of the language's PDE approach.
Such a compiler would be an example of what some authors have termed ``global-to-local (GTL)" compilation \parencite{hamann_space-time_2010}.
GTL compilation seeks to turn on its head the usual challenge of emergence: It seeks not to predict collective behavior from simple rules for individuals, but to derive such rules based on desired collective behavior.
In this section we explore a novel, embodied variation of smoothed particle hydrodynamics (SPH) swarm robotic control as a partial basis for a GTL compiler for Morphgen.

SPH itself is a meshfree Lagrangian numerical method used in physics simulations and was introduced by \textcite{gingold_smoothed_1977} and \textcite{lucy_numerical_1977}.
SPH was first proposed as a method for controlling robot swarms by \textcite{perkinson_decentralized_2005}.
In SPH robotic literature to date, researchers have used SPH to cause swarms to emulate fluids or bodies much like fluids (e.g., \cite{pimenta_swarm_2013, fujiwara_self-swarming_2014, maningo_formation_2016, silic_anisotropic_2018}).
Such use is in keeping with typical use of SPH in physics to simulate fluids.
However, we believe the true potential of SPH in robotic control lies in its ability to simulate non-physical PDEs.
A step toward this goal was taken by \textcite{pac_control_2007}, who design swarms that emulate fluids, but with non-physical variation over time in fluid parameters (see \textcite{tilki_fluid_2015} for a more recent paper continuing this work).
Our work goes further, applying SPH to Morphgen programs that do not resemble physical fluid behavior.
This is one key contribution that our work makes with respect to the SPH robotic control literature, which we address further in \Secref{compiler}.
However, we first discuss our other key contribution, a proposal for overcoming a communications challenge posed by traditional SPH robotics.

\subsection{Natural Smoothing Functions} \label{sec:nsf}
If an agent can produce a physical field around itself having certain properties, then implicit in this field is what we term a \emph{natural smoothing function} (NSF).
We will refer to such physical fields as ``NSF fields" to distinguish them from the potentially more abstract fields which NSF fields may be used to implement, as described in the next section.
The NSF field around a given agent must satisfy the following:
\begin{itemize}
    \item The strength or intensity (we will say \emph{value}) of the NSF field must decrease smoothly and monotonically with distance from the agent.
    \item The NSF field's value must approach zero with distance, and quickly enough that the integral of value over all space converges.
    \item The NSF field must be roughly symmetrical around the agent.
    \item The agent must have control over the amplitude of the NSF field independently of its shape.
\end{itemize}
Finally, the NSF fields of different agents must sum naturally in the environment, and each agent must be able to sense locally both the value and gradient of the summed NSF field.
If these requirements are met, then groups of such agents can implement SPH robotic control.
The first three of these requirements are derived from general requirements on SPH smoothing functions; see \textcite{liu_smoothed_2003} for a discussion.
Another key requirement is that the integral of the smoothing function be unity, which we address in the next section.

In later sections, we explore one possibility in depth: that of agents that release and detect dilute substances that diffuse and degrade in a surrounding aqueous medium.
This scenario is inspired by the cells of a developing embryo as it undergoes morphogenesis.
We refer to the smoothing functions implicit in this mechanism as ``morphogen smoothing functions" (MSF).
A second possibility, not discussed further in the present work, is that of agents producing and sensing electromagnetic or acoustic NSF fields distinguished by frequency.
Before addressing MSFs specifically, we discuss the NSF framework generally.

\subsection{Incorporating SPH Estimates into the NSF Framework}

In contrast to the term ``NSF field," we use the term ``field" by itself to describe more general, and potentially abstract, entities.
These are the fields which traditional SPH seeks to simulate and which SPH swarm robotic control seeks to generate.
Examples in the context of artificial morphogenesis are density, velocity, or differentiation state of a tissue or agent continuum.

A diversity of SPH-based estimates of fields and their derivatives exist. 
In general, an SPH estimate at a point is the sum over neighboring particles of a smoothing function, or one of its derivatives, multiplied by some expression involving various spatially-varying quantities.
Each of these quantities (e.g., mass, density, or velocity) is associated either with the location of the desired estimate or with the location of a given neighboring particle.
Some SPH estimates require only quantities associated with the locations of neighboring particles, and such forms permit ready incorporation into the NSF framework.
The most important example is the estimate of a field itself at location $x$:
\begin{equation} \label{eq:fsph}
    \left\langle f(x) \right\rangle = \sum_j \frac{m_j}{\rho_j}f(x_j)W_j(x),
\end{equation}
where $m_j$, $\rho_j$, and $f(x_j)$ are the mass of, local density at, and field value at, each neighboring particle $j$.

We can perform this estimate in the NSF framework if we assume that each agent can produce an NSF field as described in \Secref{nsf}.
If an agent $j$ controls the amplitude of its NSF field so that it has unity integral, then the measurement by an agent $i$ of the value of the NSF field produced by agent $j$ is equivalent to the evaluation of $W_j(x_i)$, where $W_j(x)$ is a smoothing function whose shape is determined by the physical nature of the produced NSF field.
If instead agent $j$ controls the amplitude of its NSF field so that its integral is $\frac{m_j}{\rho_j}f(x_j)$, then the measurement by agent $i$ of agent $j$'s NSF field is the entire term $\frac{m_j}{\rho_j}f(x_j)W_j(x_i)$.
Because the NSF fields of each neighbor $j$ sum naturally in the environment, the total NSF field value measured locally by agent $i$ is the desired local estimate of a field $\left\langle f(x_i) \right\rangle$.
(This procedure requires agents to estimate their local densities, which they can do through a similar NSF estimate based on $\left\langle \rho(x_i) \right\rangle = \sum_j m_j W_j(x_i)$.)

Because Eq.\ \ref{eq:fsph} is valid at points between particles as well as at particles, a spatial derivative of the field of SPH estimates is an estimate of that derivative in the corresponding field function.
For example,
\begin{equation} \label{eq:nsfgrad}
    \left\langle \nabla f(x_i) \right\rangle = \nabla_{x_i}\sum_j \frac{m_j}{\rho_j}f(x_j)W_j(x_i).
\end{equation}
In traditional SPH, where a separate calculation is needed for an SPH estimate at each point, the above fact is not directly useful.
Instead, much of the value of traditional SPH comes from the ability to rearrange expressions so that derivatives are applied only to the smoothing function and can be taken analytically.
For example,
\[
    \left\langle \nabla f(x_i) \right\rangle = \sum_j \frac{m_j}{\rho_j}f(x_j)\nabla_{x_i}W_j(x_i).
\]

By contrast, in the NSF framework, SPH estimates are physically embodied by the total NSF field at a given point.
Therefore, if technologically feasible, an agent can make an SPH estimate of a spatial derivative by sensing that derivative in the local NSF field.
In particular, we assume agents can sense NSF field gradients locally, for example using differences between multiple sensors at their boundary, such that Eq.\ \ref{eq:nsfgrad} is directly useful.

Where such direct estimates are not practical, it may be possible to manipulate other methods from the SPH literature into forms compatible with the NSF framework.
The Laplacian operator is an important example.
\textcite{huang_improved_2016} derive an SPH form of the Laplacian operator requiring no derivatives.
In two dimensions:
\[
    \left\langle\Delta f(x_i)\right\rangle = \frac{2}{\alpha} \sum_j \left(f(x_j) - f(x_i)\right) W_j(x_i) \frac{m_j}{\rho_j},
\]
where $\alpha$ is a constant that depends on an integral relating to a given smoothing function.
Although it is not possible in the NSF framework to take the pairwise difference as shown, we can rearrange and note that $\sum_j W_j(x_i) \frac{m_j}{\rho_j}$ is just an SPH estimate of unity, and so can be replaced by 1.
This gives us a form we can calculate in the NSF framework:
\begin{equation}
    \begin{aligned}
        \left\langle\Delta f(x_i)\right\rangle 
        &= \frac{2}{\alpha} \left( \sum_j \frac{m_j}{\rho_j} f(x_j) W_j(x_i) - \sum_j \frac{m_j}{\rho_j} f(x_i) W_j(x_i) \right) \\
        &= \frac{2}{\alpha} \left( \sum_j \frac{m_j}{\rho_j} f(x_j) W_j(x_i) - 1 \cdot f(x_i) \right) \\
        &= \frac{2}{\alpha} \left( \left\langle f(x_i) \right\rangle - f(x_i) \right) .
    \end{aligned}
\end{equation}

\subsection{Benefits of NSF Framework}

NSF-based SPH robotic control addresses communication challenges and inefficiencies in traditional SPH robotic control.
Traditional methods require each agent, for each SPH estimate it makes, to communicate separately with $\mathcal{O}(n)$ other agents, where $n$ is the number of agents in its neighborhood.
In contrast, NSF-based agents need only sense an NSF field value locally, a $\mathcal{O}(1)$ operation.
In establishing separate data links with its neighbors, a traditional agent must avoid communication interference with its neighbors, all of which may be simultaneously trying to establish links with \emph{their} neighbors.
By contrast, NSF-based agents communicate implicitly by sensing an NSF field locally and manipulating their contribution to it; therefore, they do not require any explicit data links or even a concept of having individual neighbors.

We can view these advantages over traditional SPH control as the benefits of embracing an embodied approach to artificial morphogenesis.
In the agent-to-agent communication of traditional SPH control, information is an abstraction atop the physical fields chosen as communication media.
Overcoming communication interference is thus a struggle against the way these fields interact in space.
By contrast, the NSF framework works with, rather than against, the interaction (as well as diminution with distance) of the physical NSF fields available to a system.
For example, morphogen-based SPH control takes advantage, in two key ways, of the behavior of morphogens in the aqueous environments typical of a growing embryo.
First, it offloads pairwise distance estimates to the natural diminution with distance of a dilute substance as it diffuses and degrades.
Second, it offloads a summation loop across many neighbors to the natural summation of dilute substances from multiple sources.

\subsection{Morphogen Smoothing Functions} \label{sec:msf}

This section explores the use of morphogens to implement NSFs as a basis for SPH control.
We refer to such NSFs as MSFs, as mentioned in \Secref{nsf}.
We first discuss the properties of these MSFs with respect to SPH generally.
Afterward we derive the morphogen production rates needed to implement SPH control using MSFs.

Throughout this section, we make several simplifying assumptions:
\begin{itemize}
    \item Each agent is a point.
    \item Each morphogen obeys Fick's law for homogeneous, isotropic diffusion (i.e., change in concentration is a constant proportion of the Laplacian).
    \item Degradation of each morphogen is uniform in the medium and proportional to morphogen concentration.
    \item Each agent is unmoving.
    \item Morphogen production, diffusion, and degradation are in equilibrium.
\end{itemize}
Later, we introduce corrections for some violations of the latter two assumptions.

For our embodied SPH approach to work, we need the SPH estimate of a field variable at a point to equal the summed concentration of an associated morphogen at that point:
\begin{equation} \label{eq:fsphc}
    \left\langle f(x) \right\rangle = \sum_j \frac{m_j}{\rho_j}f(x_j)W_j(x) = \sum_j c_j(x),
\end{equation}
where $c_j(x)$ is the morphogen concentration at $x$ attributable to neighboring agent $j$.
Each agent must therefore choose to produce this morphogen at a rate $r_j$ such that 
\begin{equation} \label{eq:cjx}
    c_j(x) = \frac{m_j}{\rho_j}f(x_j)W_j(x).
\end{equation}
Let $C_j$ be the total amount of morphogen in the environment due to agent $j$, at equilibrium:
\[ C_j = \int_\Omega c_j(x)dx. \]
If $k$ is the morphogen's decay rate, then $C_j$ is related to $r_j$ by
\begin{equation} \label{eq:Cjdyn}
    \frac{dC_j}{dt} = r_j - kC_j 
\end{equation}
and therefore reaches equilibrium when 
\begin{equation} \label{eq:Cjeq}
    C_j = \frac{r_j}{k} . 
\end{equation}
Solving Eq.\ \ref{eq:cjx} for $W_j(x)$, integrating both sides, and considering the requirement of unity for the integral of $W_j(x)$, we have
\[ \int_\Omega W_j(x)dx = \int_\Omega \frac{\rho_j}{m_j f(x_j)}c_j(x)dx = \frac{\rho_j}{m_j f(x_j)} C_j = \frac{r_j \rho_j}{k m_j f(x_j)} = 1. \]
We can now solve for $r_j$ to determine the correct production rate for agent $j$:
\begin{equation} \label{eq:rj}
    r_j = \frac{k m_j f(x_j)}{\rho_j}. 
\end{equation}
Under these simplified assumptions, the SPH smoothing functions implied by this production rate are described in Table \ref{tab:msf}.

\PreChap{}{\begin{landscape}}
    \begin{table}
        \caption{Idealized morphogen smoothing functions. The row labeled ``Physical, normalized'' refers to the physical part of the general solution to the relevant PDE, normalized so that its integral over space is unity. $r$ is the magnitude of displacement. $J_0$ and $Y_0$ are Bessel functions of the first and second kind respectively, of order zero. $K_0$ is a modified Bessel function of the second kind, of order zero. $E$ is diffusion rate and $k$ is decay rate, and in the final row we make the substitution $h = \sqrt{\frac{E}{k}}$ to obtain forms resembling the typical presentation of SPH smoothing functions.}
        \label{tab:msf}
        \makegapedcells
        $\begin{tabu}{llll}
            & \text{One dimension} & \text{Two dimensions} & \text{Three dimensions} \\ \hline
            \text{Radial form of PDE} &
            E\frac{\partial^2\rho}{\partial r^2} - k\rho = 0 &
            \frac{E}{r}\frac{\partial \rho}{\partial r} + E\frac{\partial^2\rho}{\partial r^2} - k\rho = 0 &
            2\frac{E}{r}\frac{\partial \rho}{\partial r} + E\frac{\partial^2\rho}{\partial r^2} - k\rho = 0 \\ \hline
            \text{General solution} &
            \rho(r) = c_1e^{-\sqrt{\frac{k}{E}}r} &
            \rho(r) = c_1J_0\left(ir\sqrt{\frac{k}{E}}\right) &
            \rho(r) = c_1\frac{e^{-\sqrt{\frac{k}{E}}r}}{r} \\
            &
            \;\;\;\;\;\;\;\; +\: c_2e^{\sqrt{\frac{k}{E}}r} &
            \;\;\;\;\;\;\;\; +\: c_2Y_0\left(-ir\sqrt{\frac{k}{E}}\right) & 
            \;\;\;\;\;\;\;\; +\: c_2\frac{e^{\sqrt{\frac{k}{E}}r}}{\sqrt{k}r} \\ \hline
            \text{Physical, normalized} &
            \rho(r) = \sqrt{\frac{k}{E}}e^{-\sqrt{\frac{k}{E}}r} &
            \rho(r) = \frac{k}{2\pi E}K_0\left(r\sqrt{\frac{k}{E}}\right) &
            \rho(r) = \frac{k}{4\pi E}\frac{e^{-\sqrt{\frac{k}{E}}r}}{r} \\ \hline
            \text{With } h = \sqrt{\frac{E}{k}} &
            \rho(r) = \frac{e^{-\frac{r}{h}}}{h} &
            \rho(r) = \frac{K_0\left(\frac{r}{h}\right)}{2\pi h^2} &
            \rho(r) = \frac{e^{-\frac{r}{h}}}{4 \pi h^2 r} 
            \PreChap{\\ \hline}{}
        \end{tabu}$
    \end{table}
\PreChap{}{\end{landscape}}

\subsection{Correcting for Transient Spurious Morphogen Smoothing Functions}

Traditional SPH assumes no lag time in information flow, but two factors prevent morphogens from conveying information instantaneously among agents.
The first is the time needed for production and degradation processes to reach steady state, which is a non-spatial consideration.
The second is the time needed for diffusion to spread information spatially from an agent to its neighbors.
These two factors lead to transient spurious MSFs, and we discuss each in turn.
In neither case do we fully correct for transient effects; instead, we make simplifying assumptions that allow us to correct for the most detrimental aspect of each effect.

\subsubsection{Non-equilibrium of Production and Degradation Rates}

We first address the non-spatial transient effect of morphogen production being out of equilibrium with degradation, which occurs whenever production rates have recently changed.
We make two simplifying assumptions to find a correction term for agents to add to their sensor readings to better approximate what those readings would be under equilibrium conditions.
First, we neglect diffusion time; in other words, we assume that each agent instantaneously injects morphogen into its entire neighborhood according to the distribution of the idealized MSF.
Second, we assume that the field variable values for all agents in a neighborhood are changing at the same rate $a$ so that $f(x_j, t) = at + b_j$ since time $t = 0$, and that until $t = 0$ the values were not changing and the system was in equilibrium.

Based on these assumptions, we wish to 
find a correction term $g(t)$ such that $\left\langle f(x, t) \right\rangle = \sum_j c_j(x, t) + g(t)$.
At a given time $t > 0$, from Eq.\ \ref{eq:Cjdyn} and Eq.\ \ref{eq:rj} the total mass of morphogen in the environment due to agent $j$ obeys
\[ \frac{dC_j(t)}{dt} = r_j(t) - kC_j(t) = \frac{k m_j (at + b_j)}{\rho_j} - kC_j(t). \]
This equation is solved by
\begin{equation} \label{eq:Cjtgen}
    C_j(t) = \alpha e^{-kt} + \frac{m_j}{\rho_j}\left( at + b_j - \frac{a}{k} \right).
\end{equation}
Because the system was in equilibrium at $t = 0$, we can substitute from Eq.\ \ref{eq:Cjeq} and Eq.\ \ref{eq:rj}: 
\[ C_j(0) = \frac{r_j(0)}{k} = \frac{m_j f(x_j, 0)}{\rho_j} = \frac{m_j b_j}{\rho_j}. \]
After substituting into Eq.\ \ref{eq:Cjtgen} and solving for $\alpha$, we have:
\[
    C_j(t) = \frac{m_j a}{\rho_j k} e^{-kt} + \frac{m_j}{\rho_j}\left( at + b_j - \frac{a}{k} \right) 
    = \frac{m_j a}{\rho_j k} e^{-kt} + \frac{m_j f(x_j, t)}{\rho_j} - \frac{m_j a}{\rho_j k},
\]
which we can solve for $f(x_j, t)$:
\begin{equation} \label{eq:fjcorr}
    f(x_j, t) = \frac{\rho_j}{m_j}C_j(t) + \frac{a}{k}\left(1 - e^{-kt}\right).
\end{equation}
Substituing Eq.\ \ref{eq:fjcorr} into Eq.\ \ref{eq:fsph} and rearranging, a correct SPH estimate of $f(x, t)$ under our non-equilibrium assumptions would therefore be
\begin{equation} \label{eq:sphfcorr}
    \left\langle f(x, t) \right\rangle 
    = \sum_j C_j(t)W_j(x) + \sum_j \frac{m_j}{\rho_j} \frac{a}{k} \left( 1 - e^{-kt} \right) W_j(x)
\end{equation}
The left summation of Eq.\ \ref{eq:sphfcorr} is the SPH estimate of $f(x, t)$ under the equilibrium assumption and is also $\sum_j c_j(x, t)$, the sensed morphogen concentration at $x$ (because we derived our production rate under the equilibrium assumption).
The right summation of Eq.\ \ref{eq:sphfcorr} is just the SPH estimate of $\frac{a}{k} \left( 1 - e^{-kt} \right)$, but this does not depend on $j$, so no SPH estimate is necessary for this term:
\[
    \left\langle f(x, t) \right\rangle
    = \sum_j C_j(t)W_j(x) + \frac{a}{k} \left( 1 - e^{-kt} \right).
\]
Thus we have our desired correction term:
\begin{equation} \label{eq:g}
    g(t) = \frac{a}{k} \left( 1 - e^{-kt} \right).
\end{equation}

We wish for each agent to keep track of $g(t)$, which is possible so long as it maintains an estimate of $\frac{\partial f(x, t)}{\partial t}$. 
(In our simulation we simply have each agent remember one previous value of $f(x, t)$ which it can divide by a known timestep.)
To see this, we first take a time derivative: $\frac{d g(t)}{d t} = ae^{-kt}$.
Note that we can rearrange Eq.\ \ref{eq:g} to get $e^{-kt} = 1 - \frac{k}{a}g(t)$.
We substitute this into the derivative along with $a = \frac{\partial f(x, t)}{\partial t}$ to obtain
\begin{equation} \label{eq:dg}
    \frac {d g(t)}{d t} = \frac{\partial f(x, t)}{\partial t} - kg(t),
\end{equation}
which implies the needed update rule for agents to follow (in our simulation, $\Delta g(t) = \Delta f(x, t) - kg(t)\Delta t$).

\subsubsection{Transient Effects of Diffusion Rate}

The time it takes for morphogen diffusion to convey information spatially also leads to violations of steady-state assumptions whenever the position or production rate of an agent has recently changed.
Here we address only transient effects due to agent motion, because in our simulations it appears to be more detrimental than spatial effects due to production rate changes.
As an agent moves, its contribution to the total morphogen field becomes stretched behind and compressed in front, with more complex distortions if the agent's path curves.
Even more complex effects arise as a result of the particularities of an agent's instrument positions and the way a given medium flows around an agent's body.
Because such particularities imply implementation-dependence of distortions, we do not pursue analytic corrections.
Instead, we propose that agents be calibrated to correct for some of this distortion, an approach we test in simulation.

The most damaging aspect of this distortion is the effect that an agent's own contribution to the morphogen field has on its gradient estimates.
In addition, an agent has no knowledge of its neighbors' individual positions and velocities, limiting its ability to correct for distortions in neighbors' contributions to the morphogen field.
For these reasons we focus calibration on correcting for distortions in an agent's own contributions to its gradient estimates.

The central idea of calibration is to have isolated agents move at constant velocities and, once steady-state relative to the agent is reached, to record the measured gradient.
A table of gradients for different speeds can be created in this way, and during operation these gradients (with appropriate interpolation) can be subtracted from measured gradients during operation to partially correct for motion-related asymmetry in the MSF.
For our simulation, we repeat this procedure with different headings to average out numerical effects related to the finite difference method grid used for diffusion.

\subsection{Beyond Fluid Dynamics: SPH as a Partial Morphgen Compiler} \label{sec:compiler}

SPH robotic control generally, and our NSF variant in particular, can be viewed as a compilation technique for Morphgen code.
The utility of SPH as a compilation technique is limited in the present work by the requirement that swarm density fields must obey the mass continuity equation $\frac{\partial \rho}{\partial t} = -\nabla \cdot (\rho \boldsymbol{u})$.
This trivially holds when swarm density is specified in Morphgen implicitly by velocity or acceleration fields, but may not hold when swarm density fields are specified directly.
For example, a nonconservative equation such as $\frac{\partial \rho}{\partial t} = 5$ cannot yet be compiled with our method.
Other fields representing swarm state are not limited in this way.
In future work we plan to relax this requirement through agent recruitment and removal or through cell-like division and apoptosis.

SPH compilation introduces a layer of abstraction or virtualization between what we might think of as Morphgen ``application" code and what we envision as a library of physical morphogens, which can also be described in Morphgen.
Consider a Morphgen program that specifies a velocity or acceleration field for a swarm, along with several other fields.
These other fields may be thought of as representing tissue differentiation, gene expression, or even morphogens, as seen in earlier examples.
To compile such a program into agent code, all non-swarm fields are treated as agent state, and a unique physical ``library" morphogen is associated with each.
(In practice, a programmer might find it convenient to specify these associations, because physical morphogens with different ratios of diffusion to degradation rates might prove more robust for implementing different fields.
However, such low-level associations are not essential in principle: Eq.\ \ref{eq:rj} specifies the correct production rate regardless of diffusion and degradation rates.)
As described in \Secref{msf}, these physical morphogens implicitly provide the smoothing functions needed for each agent to make SPH estimates of a given field and its derivatives.

To further illustrate this layer of abstraction, consider a Morphgen application program that specifies what the programmer conceptualizes as a morphogen (for example, if the Morphgen code specifies a diffusion and degradation rate as well as sources and sinks).
The SPH compilation process would nonetheless treat this field as a swarm state variable and associate it with a physical morphogen, whose physical parameters are theoretically independent of the abstract morphogen described by the application code.
This layer of abstraction frees the programmer from the need to find a physical substance with particular parameters.
Instead, we envision engineering a small library of physical morphogens that can be repurposed to implement a variety of fields in different Morphgen programs.

As mentioned in \Secref{toward}, SPH swarm robotic control has focused on enabling swarms to behave somewhat like fluids, as described by the Navier-Stokes or Euler equations.
Swarms have therefore been specified with acceleration fields, and other fields have been designed to simulate such physical quantities as energy and pressure.
In the context of our current work, we prefer to reframe these applications as special cases of SPH compilation of implicit Morphgen-like programs that happen to describe roughly physical fluid-like behavior.
(This is a simplification; for example, \textcite{pimenta_swarm_2013} derive one force using SPH, and add to it a body force derived separately from a potential field.)

In comparison to past fluid-related SPH swarm control, the example that follows illustrates some of this potential generality of SPH compilation.
Rather than specifying physics-inspired quantities such as energy or pressure, arbitrary fields useful in solving a path-finding problem are specified.
And rather than specifying acceleration, velocity is instead specified, making swarm behavior more similar to flocking than to Newtonian dynamics. 
(Velocity-based motion is also a natural fit for fluid media with low Reynolds numbers such that propulsion force is approximately proportional to velocity rather than to acceleration.
However, agents could integrate or differentiate on their own to achieve, respectively, acceleration-based control where force is proportional to velocity or velocity-based control where force is proportional to acceleration.)

\subsection{Example: Path-finding}

To illustrate our method, we simulate a two-dimensional environment that is inspired by, but does not rigorously implement, an aqueous environment with low Reynolds number, as would be found in a developing embryo.
Agents are modeled as discs.
Sensors are placed around the surface at 90\textdegree\ intervals, with morphogen production instruments on the surface halfway between each pair of sensors.
Agents have full control over their velocity, but no propulsion mechanism is explicitly modeled; agents' locations are simply changed.
As agents move, the surrounding medium and morphogens flow around the agents' bodies.
This flow is qualitatively similar to laminar flow but does not strictly obey fluid dynamics.
Motion of the aqueous medium and morphogens is calculated using a finite difference method.
Some random noise is applied to agents' motion to qualitatively simulate Brownian motion, and some bias and noise is applied to sensor readings.

In the Morphgen program below, $\langle T \rangle$ refers to the SPH estimate of $T$, as opposed to an agent's internal stored value.
\begin{align*}
\word{substance}\ \mbox{swarm}:\\
  \word{scalar\ fields:}\\
  \rho &  \mbox{\ \ \ \ // swarm density} \\
  A &  \mbox{\ \ \ \ // environmental cue for lower-left square} \\
  B &  \mbox{\ \ \ \ // environmental cue for upper-right square} \\
  S &   \mbox{\ \ \ \ // to diffuse from lower-left square} \\
  \displaybreak[0] 
  T &   \mbox{\ \ \ \ // to follow gradient of $S$ back from upper-right square} \\
\word{vector\ fields:} \\
  \vect{V} &   \mbox{\ \ \ \ // swarm velocity} \\
  \displaybreak[0] 
\ \ \\
   \word{behavior:\ \ \ \ }\\
     \word{params:}\\
       \theta_\rho &= 0.02 \\
       \theta_A &= 0.2 \\
       \theta_B &= 0.2 \\
       \theta_{T_1} &= 0.1 \\
       \displaybreak[0] 
       \theta_{T_2} &= 0.05 \\
       \kappa_{S_1} &= 500 \\
       \kappa_{S_2} &= 500 \\
       \kappa_{S_3} &= 1 \\
       \displaybreak[0] 
       \kappa_{S_4} &= 100 \\
       \kappa_{T_1} &= 500 \\
       \kappa_{T_2} &= 600 \\
       \kappa_{T_3} &= 0.002 \\
       \displaybreak[0] 
       \kappa_{T_4} &= 60 \\
       \kappa_V &= 0.0003 \\
       \displaybreak[0] 
       \kappa_{V_\text{max}} &= 0.003 \\
       \word{let}&\ \phi = \arccos\left(\frac{\nabla S \cdot \nabla T}{\|\nabla S\| \|\nabla T\|}\right) \\
       \Change S &\decrby [\rho < \theta_\rho \lor B > \theta_B]\kappa_{S_1} S\\
       \Change S &\incrby [A > \theta_A \vee S>1]\kappa_{S_2}(1-S) \\
       \displaybreak[0] 
       \Change S &\incrby [A \le \theta_A]\left(\kappa_{S_3}\nabla^2 S - \kappa_{S_4}S\right) \\
       \Change T &\incrby [B > \theta_B]\kappa_{T_1}(1 - T) \\
       \Change T &\incrby [\|\nabla S\| > 0 \land \|\nabla T\| > 0 \land \langle T \rangle > \theta_{T_1}]\left(1 - \frac{\phi}{\pi}\right)^6 \kappa_{T_2} \\
       \Change T &\incrby \kappa_{T_3}\nabla^2 T \\
       \Change T &\decrby \kappa_{T_4}T^2 \\
       \displaybreak[0] 
       \Change T &\incrby (\vect{V} \cdot \nabla T) [\langle T \rangle > \theta_{T_2} \land T < \theta_{T_1}] \\
       \word{let}&\ \vect{V_0} = \frac{1}{\rho} \kappa_V 2T \ln \left(\frac{8\rho}{T}\right)\left(\frac{\nabla \rho}{T} - \frac{\rho \nabla T}{T^2} \right) \\
       \vect{V} &= \kappa_{V_\text{max}} \tanh \left( \frac{\|\vect{V_0}\|}{\kappa_{V_\text{max}}}\right) \frac{\vect{V_0}}{\|\vect{V_0}\|}
    \end{align*}

\begin{table}
    \caption{Path-finding example.
    Agents use SPH to establish path between yellow and turquoise squares.}
    \label{tab:path}
    \begin{tabu}{cccc}
        &
        166 agents, &
        1186 agents, &
        10675 agents, \\
        &
        diameter 0.016 &
        diameter 0.006 &
        diameter 0.002 \\
        \subfloat{Time = 0} &
        \subfloat{\includegraphics[width=0.30\linewidth]{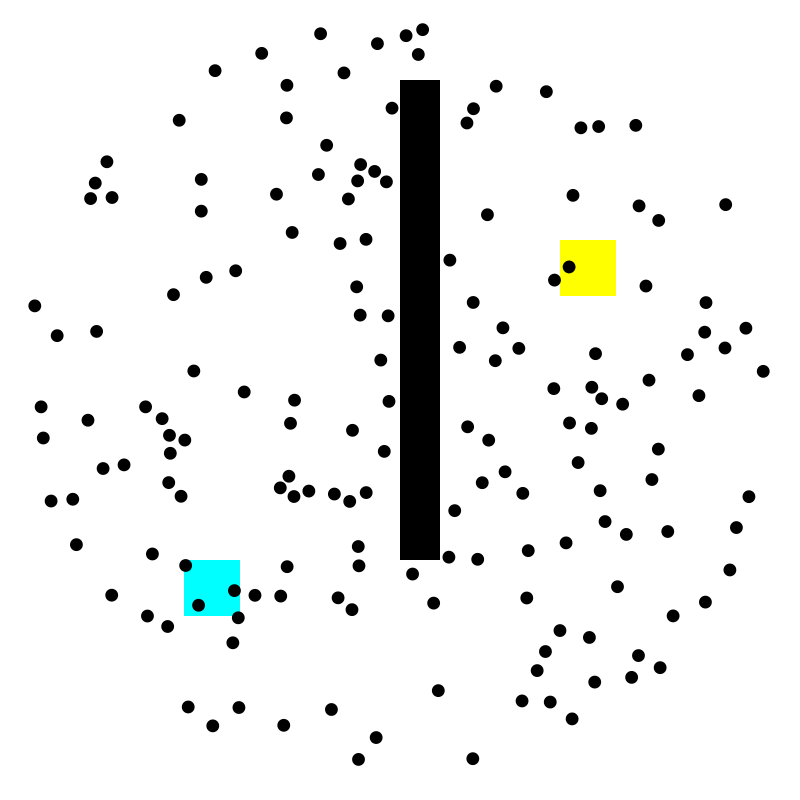}} &
        \subfloat{\includegraphics[width=0.30\linewidth]{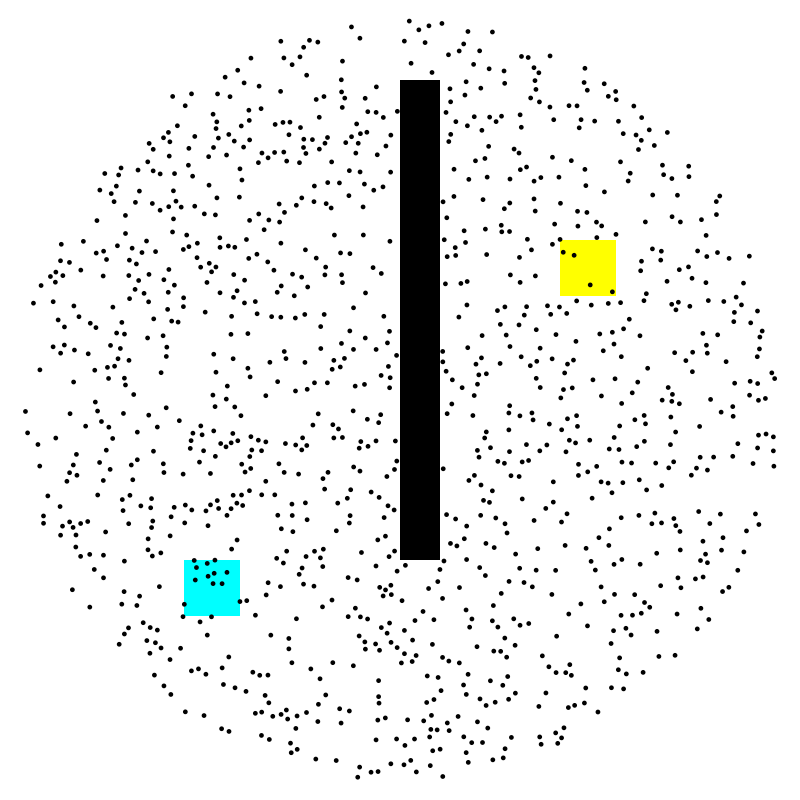}} &
        \subfloat{\includegraphics[width=0.30\linewidth]{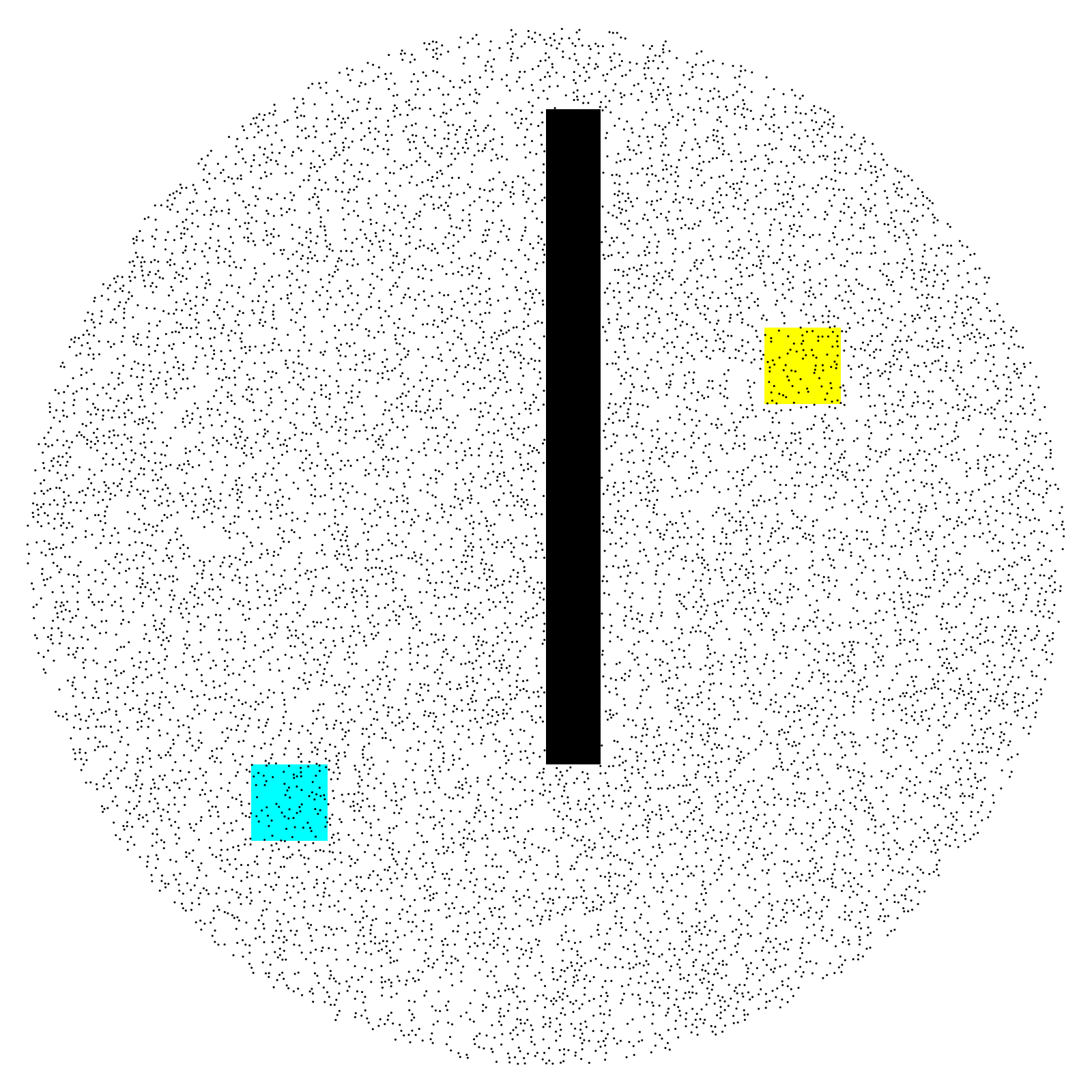}} \\
        \subfloat{Time = 20} &
        \subfloat{\includegraphics[width=0.30\linewidth]{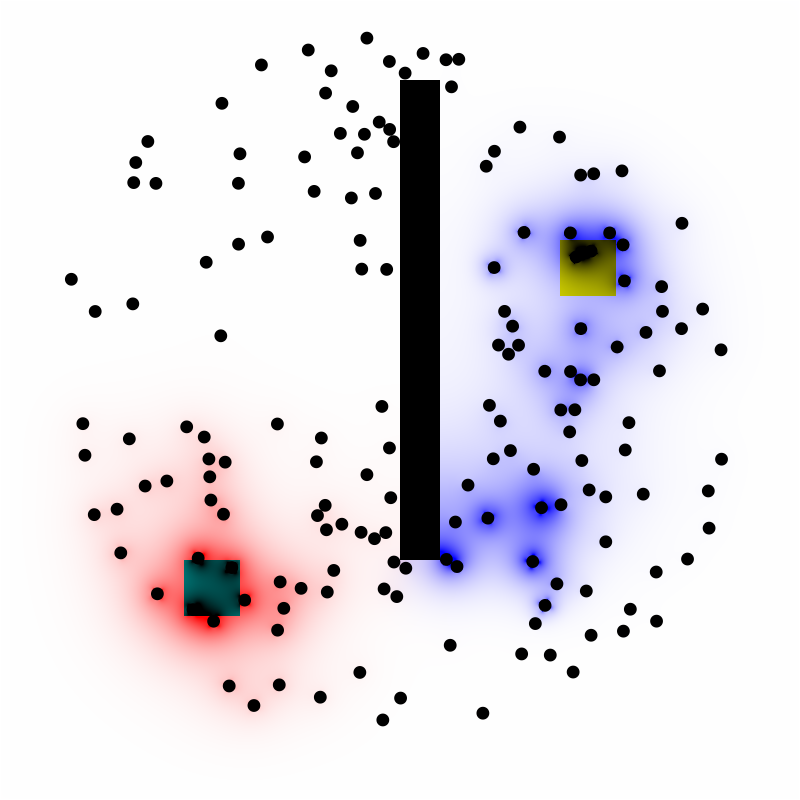}} &
        \subfloat{\includegraphics[width=0.30\linewidth]{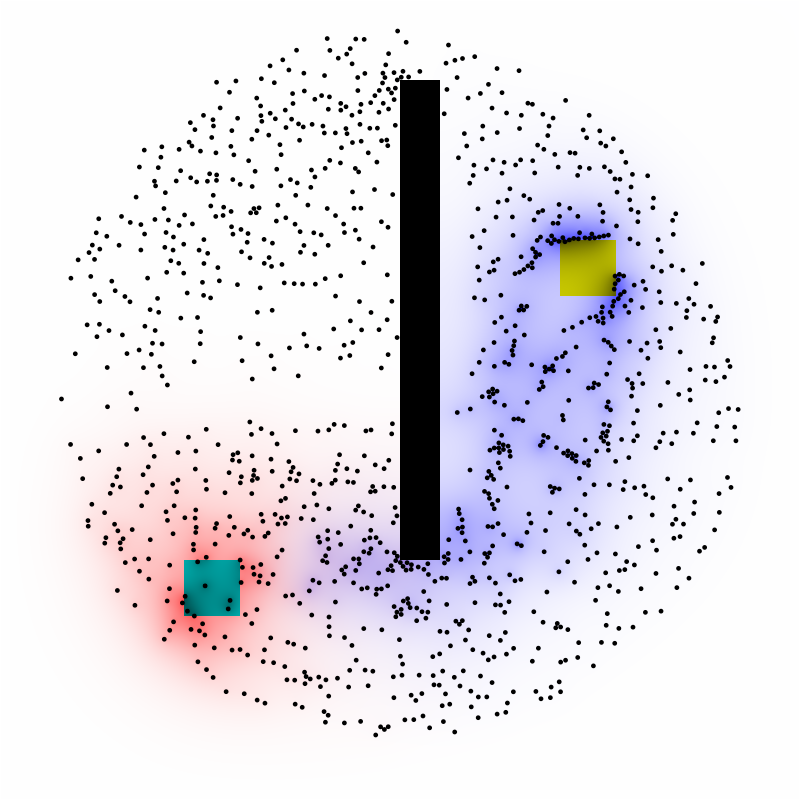}} &
        \subfloat{\includegraphics[width=0.30\linewidth]{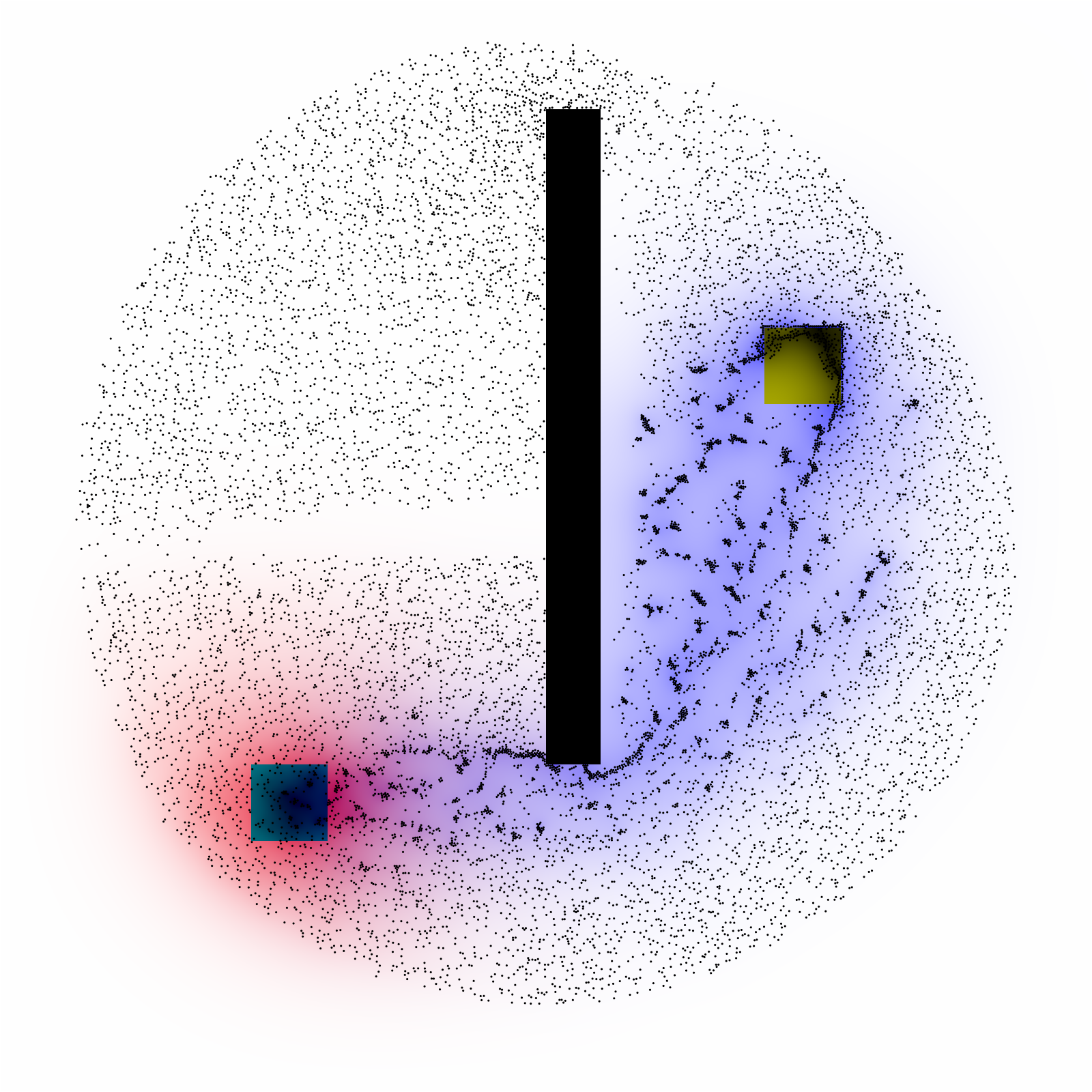}} \\
        \subfloat{Time = 70} &
        \subfloat{\includegraphics[width=0.30\linewidth]{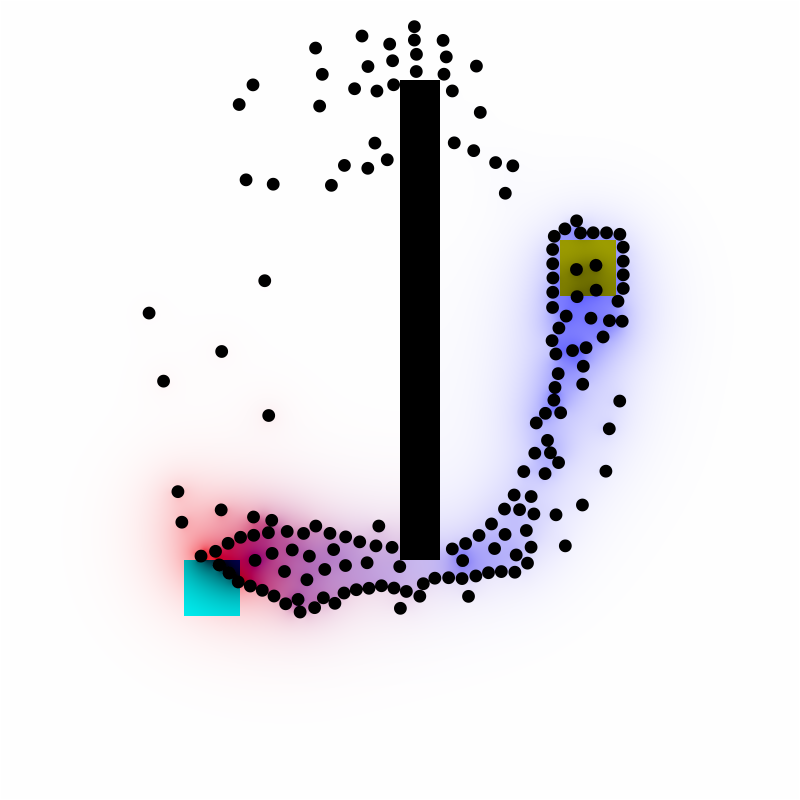}} &
        \subfloat{\includegraphics[width=0.30\linewidth]{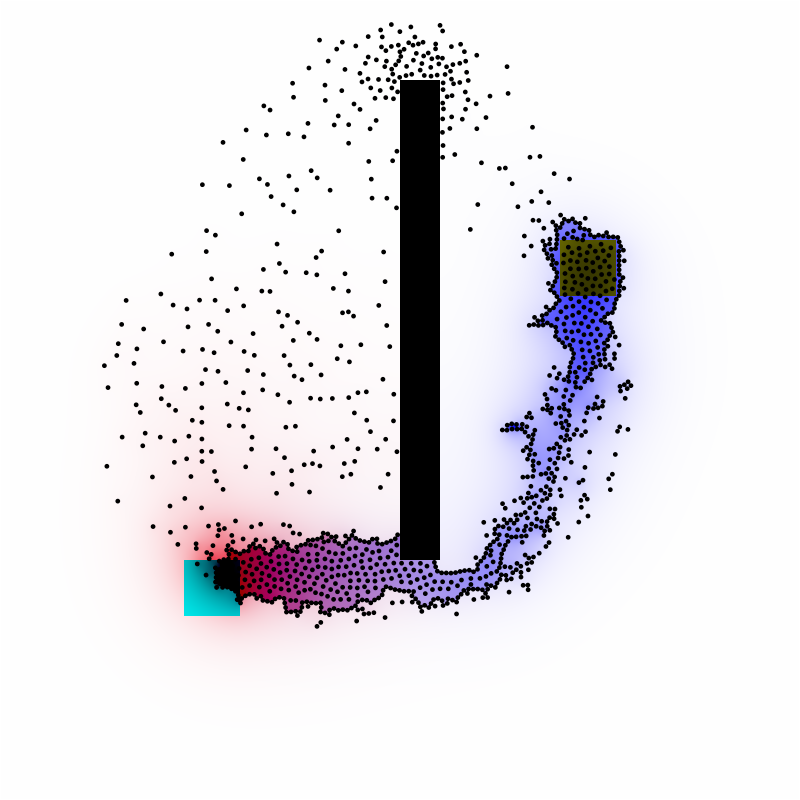}} &
        \subfloat{\includegraphics[width=0.30\linewidth]{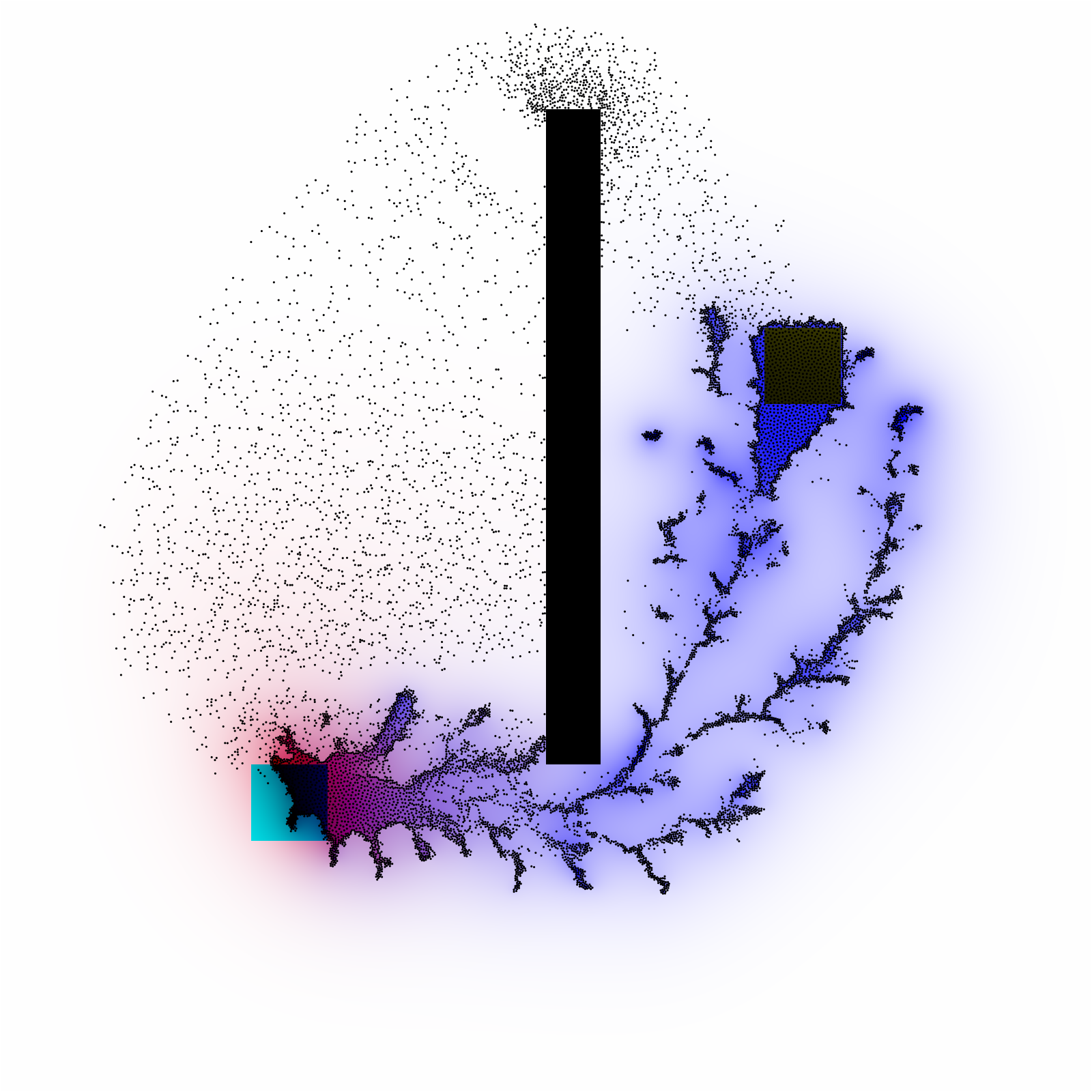}} \\
        \subfloat{Time = 270} &
        \subfloat{\includegraphics[width=0.30\linewidth]{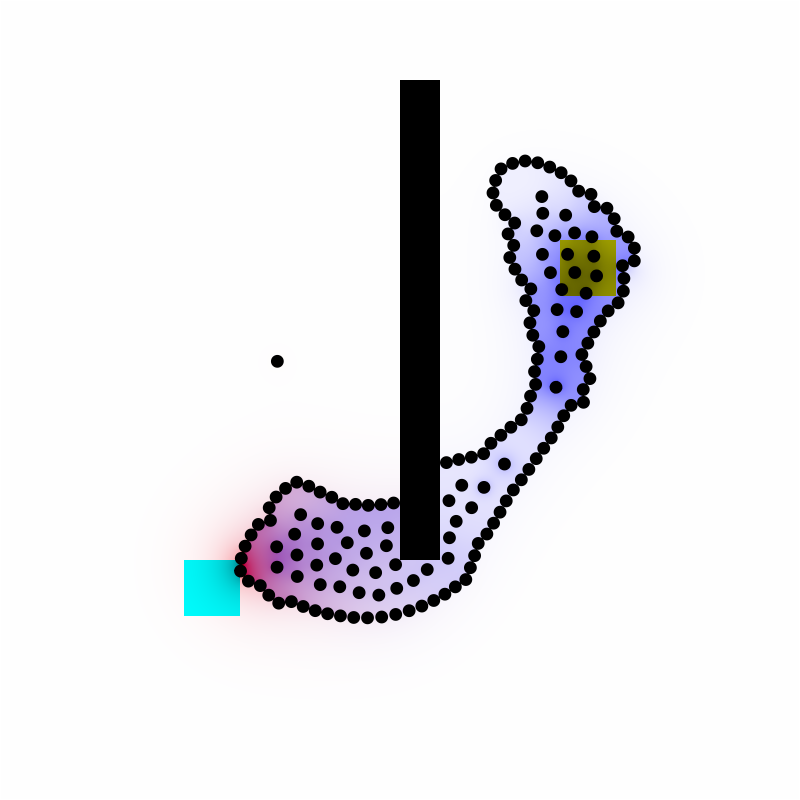}} &
        \subfloat{\includegraphics[width=0.30\linewidth]{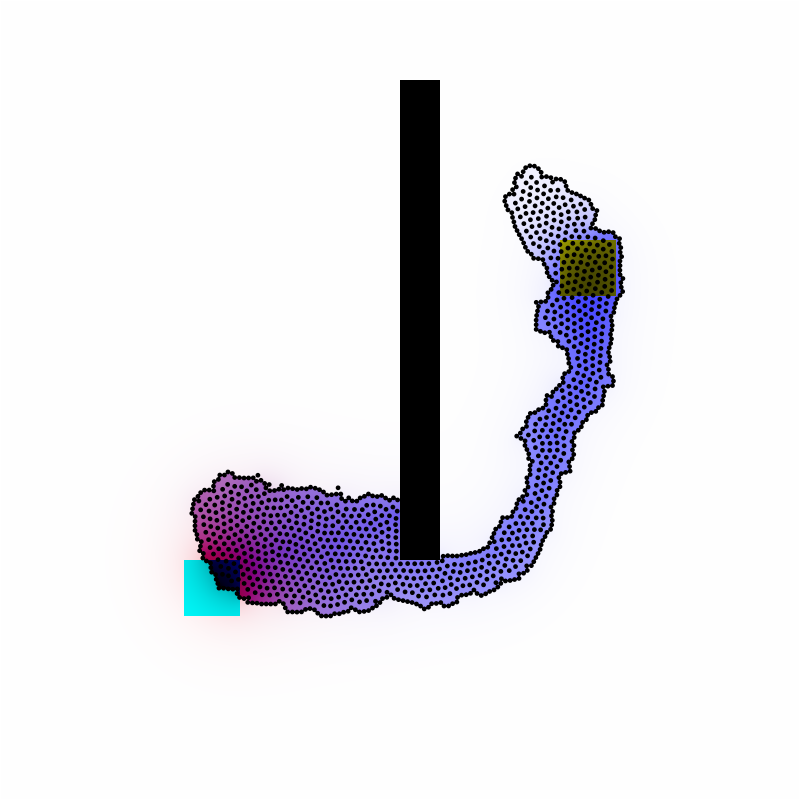}} &
        \subfloat{\includegraphics[width=0.30\linewidth]{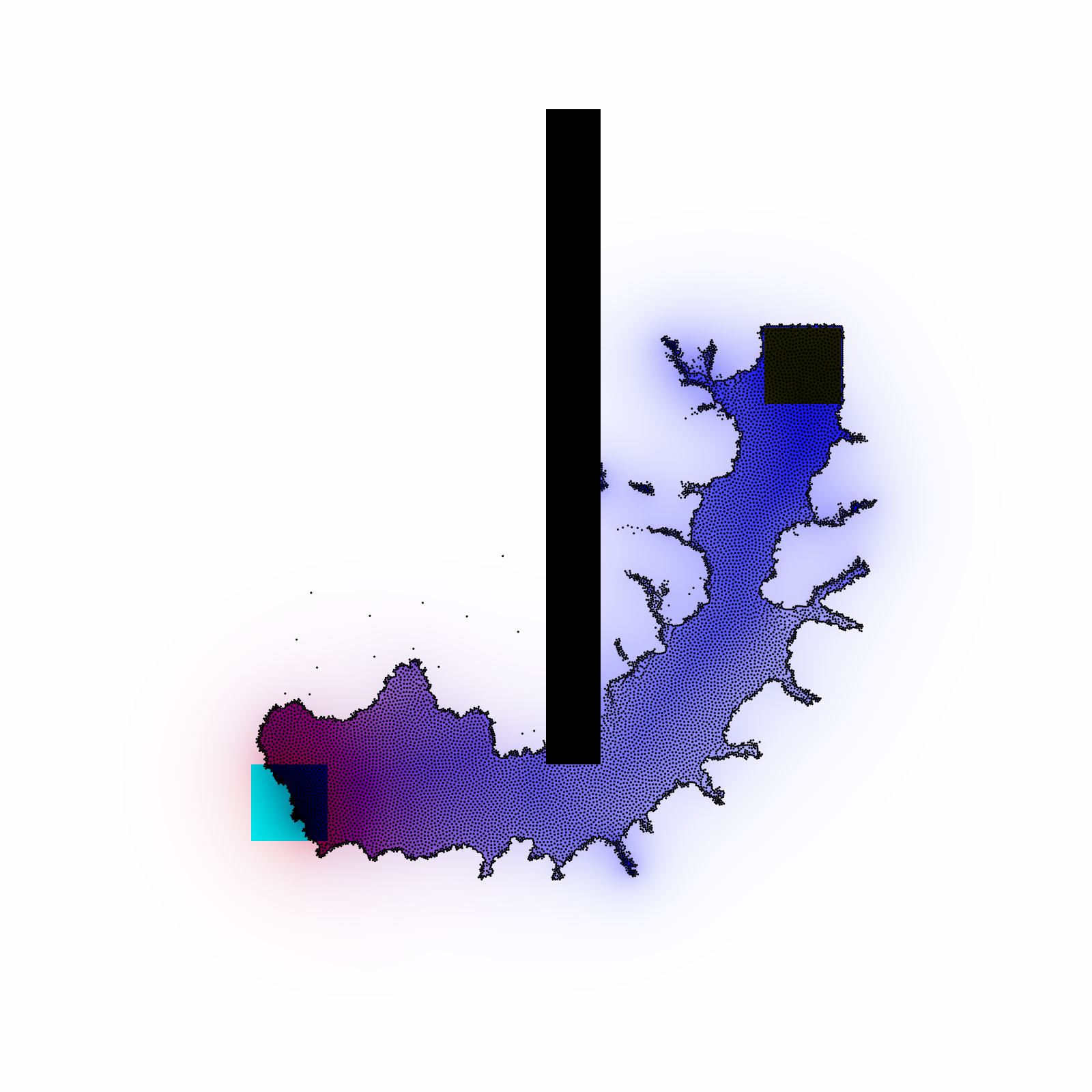}} \\
    \end{tabu}
\end{table}

Simulation of the above Morphgen program is shown in Table \ref{tab:path}.
All PDEs and simulation parameters, including starting density, are the same for all three runs shown, with only agent diameter and number varying.
To hold starting density roughly constant across the eightfold difference in agent diameter, agent number varies about 64-fold.
The qualitative similarity across this range demonstrates the potential of Morphgen and SPH to facilitate scalable swarm dynamics.

We describe some of the key components of this program, which aligns agents along the shortest path between two distinct environmental cues.
Agents that sense a cue $A$ represented by the lower-left square set a high value for the variable $S$.
Other agents implement a diffusion-decay equation to establish the field distribution of $S$.
(This equation is independent of the diffusion-decay equation that models the physical behavior of the morphogen that implements $S$ in the SPH scheme.)
Agents that sense a second cue $B$ represented by the upper-right square as well as a non-zero gradient of $S$ set a high value for variable $T$.
Other agents increase their value of $T$ depending on how nearly opposite are the gradient directions of $S$ and $T$ and whether a neighborhood estimate of $T$ is above a threshold.
Otherwise, $T$ diffuses slowly.
Agents follow the gradient of a function involving both $T$ and local density such that a target density is achieved where $T$ is high.
Agents' speed saturates with the $\tanh$ function so that dynamics are similar to flocking behavior when far from optimum arrangement but approach rest when closer to that optimum.

Because of the Lagrangian nature of SPH, field variables are naturally carried along with agents, which may or may not be desired.
At low speeds and with field dynamics that are somehow (as in this program) tethered to environmental cues, the distinction is often subtle in practice.
When it is desired that Morphgen PDEs refer to an Eulerian frame, advection can be added counter to agent velocity, as reflected by $\Change T \incrby \vect{V} \cdot \nabla T$ in the program above.
In this program, we only add this counter-advection where it helps the growing $T$ region to continue its growth against the opposing motion of agents following the $T$ gradient.

\section{Conclusions}
We have argued that embryological morphogenesis provides a model of how massive swarms of microscopic agents can be coordinated to assemble complex, multiscale hierarchical structures, that is, artificial morphogenesis or morphogenetic engineering.
This is accomplished by understanding natural morphogenetic processes in mathematical terms, abstracting from the biological specifics, and implementing these mathematical principles in artificial systems.

As have embryologists, we have found partial differential equations and continuum mechanics to be powerful mathematical tools for describing the behavior of very large numbers of very small agents, in fact, taking them to the continuum limit.
In this way we intend to have algorithms that scale to very large swarms of microrobots.

To this end we have developed a PDE-based notation for artificial morphogenesis and designed a prototype morphogenetic programming language.
This language permits the precise description of morphogenetic algorithms and their automatic translation to simulation software, so that morphogenetic processes can be investigated.

We illustrated the morphogenetic programming language and morphogenetic programming techniques with two examples.
The first addressed the problem of routing dense bundles of many fibers between specified regions of an artificial brain.
Inspired by axonal routing during embryological development, we used a modified flocking algorithm to route fiber bundles between origins and destinations while avoiding other bundles.
Simulations showed that this algorithm scaled over at least four orders of magnitude, with swarms of 5000 agents.
Then we took the number and size of agents to the continuum limit and showed how morphogenetic programming could be used to coordinate a massive swarm of agents to lay down a path between designated termini while avoiding obstacles.

Our second example showed how a natural morphogenetic model---the clock-and-wavefront model of spinal segmentation---could be applied in morphogenetic engineering both in a similar context---creating the segmented ``spine'' of an insect-like robot body---and for a different purpose---assembling segmented legs on the robot's spine.
A massive swarm of microscopic agents is supplied from external sources and guided to the assembly sites, where they begin a process of differentiation coordinated by the emission of and response to several morphogens.
As demonstrated in simulation, this reasonably complex process can be controlled to assemble a structure with a spine and legs with specified numbers and sizes of segments.

Finally, we showed how an embodied variation of smoothed particle hydrodynamics (SPH) swarm robotic control can be applied to the global-to-local compilation problem, that is, the derivation of individual agent control from global PDE specifications.
By these means, available physical morphogens can be used to implement the abstract morphogens and other fields required for a morphogenetic process. 
The physical morphogens define natural smoothing functions that can be used in SPH to estimate the concentrations, gradients, and other functions of morphogenetic fields.

\PreChap{\newpage}{} 
\bibliographystyle{plain}
\bibliography{BJM2019,ECrefs,allen}

\end{document}